\documentclass[letterpaper,12pt]{article}
\pdfoutput=1
\usepackage{jheppub}
\usepackage{amsmath}
\usepackage{graphicx}
\usepackage{subfig}

\def\({\left(}
\def\){\right)}
\def\[{\left[}
\def\]{\right]}
\def\ph1{\phantom{1}}

\newcommand{\labell}[1]{\label{#1}}

\newcommand{\reef}[1]{(\ref{#1})}

\newcommand{\mt}[1]{\textrm{\tiny #1}}

\newcommand{\be}{\begin{equation}}
\newcommand{\ee}{\end{equation}}
\newcommand{\ba}{\begin{aligned}}
\newcommand{\ea}{\end{aligned}}
\newcommand{\beq}{\begin{equation}}
\newcommand{\eeq}{\end{equation}}
\newcommand{\beqa}{\begin{eqnarray}}
\newcommand{\eeqa}{\end{eqnarray}}
\newcommand{\beqar}{\begin{eqnarray*}}
\newcommand{\eeqar}{\end{eqnarray*}}
\newcommand{\eg}{{\it e.g.,}\ }
\newcommand{\ie}{{\it i.e.,}\ }

\newcommand{\A}{\mathcal{A}}

\newcommand{\cL}{\mathcal{L}}

\newcommand{\R}{\mathcal{R}}

\newcommand{\al}{\alpha}
\newcommand{\bt}{\beta}
\newcommand{\ga}{\gamma}
\newcommand{\de}{\delta}
\newcommand{\la}{\lambda}

\newcommand{\Gn}{G_\mt{N}}
\newcommand{\hS}{{\hat S}}
\newcommand{\lp}{\ell_{\mt{P}}}

\newcommand{\sbh}{S_\mt{BH}}
\newcommand{\name}{differential\ } 

\preprint{KUNS-2486}

\title{Holographic Holes in Higher Dimensions}

\author[a]{Robert C. Myers,}
\author[a,b]{Junjie Rao}
\author[a,c]{and Sotaro Sugishita}

\affiliation[a]{Perimeter Institute for Theoretical Physics, Waterloo, Ontario N2L 2Y5, Canada}
\affiliation[b]{Department of Physics \& Astronomy and Guelph-Waterloo Physics Institute,\\
University of Waterloo, Waterloo, Ontario N2L 3G1, Canada}
\affiliation[c]{Department of Physics, Kyoto University, Kyoto 606-8502, Japan}

\emailAdd{rmyers@perimeterinstitute.ca}
\emailAdd{jrao@perimeterinstitute.ca}
\emailAdd{sotaro@gauge.scphys.kyoto-u.ac.jp}

\abstract{We extend the holographic construction of \cite{hole} from
AdS$_3$ to higher dimensions. In particular, we show that the Bekenstein-Hawking
entropy of codimension-two surfaces in the bulk
with planar symmetry can be evaluated in terms of the
`\name entropy' in the boundary theory. The \name entropy is a certain quantity
constructed from the entanglement entropies associated with a family of regions
covering a Cauchy surface in the boundary geometry. We demonstrate that
a similar construction based on causal holographic information fails in higher
dimensions, as it typically yields divergent results. We also show that
our construction
extends to holographic backgrounds other than AdS spacetime and can
accommodate Lovelock theories of higher curvature gravity.}

\begin{document}
\maketitle

\section{Introduction}

Remarkably, the entropy of a black hole is embodied in the spacetime geometry, as
expressed by the Bekenstein-Hawking (BH) formula \cite{beks,BH:1973}:
\be \sbh= \frac{{\cal A}}{4\Gn} \,, \labell{prop0} \ee
where $\cal A$ is the area of (a cross-section of) the event horizon. In fact, this expression
applies equally well to any Killing horizon \cite{Jacobson:2003wv}, including de Sitter \cite{DS} and
Rindler \cite{ray} horizons, as well as horizons in
higher dimensions.\footnote{In $d$ spacetime dimensions,  we are using `area' in a
generalized sense here to denote the volume of a spatial codimension-two subspace,
\ie the `area' has units of $\text{\emph{length}}^{d-2}$.}
Further, this expression \reef{prop0} extends to a
more general geometric formula, the `Wald entropy', to describe the horizon entropy in
gravitational theories with higher curvature interactions \cite{WaldEnt}.

Recently, it was proposed that the above expression
\reef{prop0} has much wider applicability and serves as a characteristic
signature for the emergence of a semiclassical spacetime geometry in a theory of quantum
gravity \cite{new1}. More precisely, the spacetime entanglement conjecture of \cite{new1}
may be stated as follows:
In a theory of quantum gravity, for any sufficiently large
region in a smooth background spacetime, one may consider the
entanglement entropy between the degrees of freedom describing
the given region with those describing its complement. First, ref.~\cite{new1}
conjectures that in this context, the contribution
describing the short-range entanglement will be finite and have a local geometric
description in terms of the geometry at the entangling surface. Further, the leading contribution
from this short-range entanglement will be given precisely by the BH formula \reef{prop0}.
Of course, an implicit assumption is that the usual Einstein-Hilbert action
(as well as, possibly, a cosmological constant term) emerges as the leading
contribution to the low energy effective gravitational action. As demonstrated in \cite{misha9},
higher curvature corrections to the gravitational action will also control the subleading contributions
to this entanglement entropy, which take a form similar to those in the Wald entropy.


One simple observation giving support for this spacetime entanglement
conjecture comes from gauge/gravity duality. In their seminal work \cite{rt1},
Ryu and Takayanagi conjectured a simple and elegant prescription
for a holographic calculation of entanglement entropy in the boundary
theory --- see also
\cite{rt2,rt3}. In particular, the entanglement entropy for a specified
spatial region $A$ in the boundary and its complement is evaluated with
 \be
S(A) = \mathrel{\mathop {\rm
ext}_{\scriptscriptstyle{a\sim A}} {}\!\!} \[\frac{{\cal A}(a)}{4\Gn}\]
 \labell{define}
 \ee
where $a\sim A$ indicates that the bulk surface $a$ is homologous
to the boundary region $A$ \cite{head,fur06a}. The symbol `ext'
indicates that one should extremize the area over all such surfaces $a$.
This prescription applies where the bulk
is described by classical Einstein gravity and was recently proved
for static backgrounds in \cite{aitor}.
Hence in this context, we are evaluating the Bekenstein-Hawking formula
\reef{prop0} on surfaces which generally do not correspond to a horizon in the
bulk.\footnote{The special case of a spherical entangling surface on the boundary
of AdS space is an exception to this general rule \cite{chm,eom1}. That is, in this
case, the extremal bulk surface corresponds to the bifurcation surface of a Rindler-like
horizon in the AdS bulk.} Further the usual bulk/boundary
dictionary equates an entropy on the boundary theory to an entropy in the bulk theory
and hence from these
holographic calculations, we can infer that the Bekenstein-Hawking formula
in eq.~\reef{define} literally yields an entropy
for the corresponding bulk surface $a$.

One may note, however, that the prescription for holographic
entanglement entropy picks out a special class of bulk surfaces, \ie extremal surfaces
with a specified set of asymptotic conditions at
the boundary of the bulk geometry. In contrast, the spacetime entanglement conjecture maintains that
the BH formula would determine the entropy associated with any such surface, whether or not it is
extremal, as well as for closed surfaces that do not reach the asymptotic boundary.
However, there is no contradiction here. The entanglement entropy in the boundary theory
has a unique value once the entangling surface and the state are specified and hence
the holographic prescription would be incomplete without specifying a specific bulk
surface on which to evaluate eq.~\reef{prop0}. We may also add that there have also been some
earlier discussions that more general bulk surfaces  may also give some entropic measure
of correlations in the boundary theory \cite{mark1,causal0}.

Recently, ref.~\cite{hole} studied whether a precise meaning could
be given to the spacetime entanglement conjecture in a more general context in the AdS/CFT
correspondence. In particular, this paper investigated whether the entropy $\sbh=
{\cal A}/4\Gn$ for closed curves in the bulk of AdS$_3$ could appear as an
observable in the two-dimensional boundary CFT. Strong sub-additivity was used
to argue that this quantity should be bounded by the following combination
of entanglement entropies
 \be
E=\sum_{k=1}^n \[\, S(I_k)-S(I_k\cap I_{k+1})\,\]
\,, \labell{residue}
 \ee
where the intervals $I_k$ cover a time slice in the boundary. In fact, it was shown
that applying the holographic prescription \reef{define} in a particular continuum limit
leads to the saturation of this bound with $E=\sbh$ ---
we review the details of their construction in section \ref{ads3}. They suggested that
$E$ corresponds to the `residual' entropy which measures the uncertainty in the
density matrix of the global state if one tries to reconstruct the density matrix from
observations of an infinite family of observers making observations in the causal
development of each interval. As the expression in eq.~\reef{residue} will be central
to our discussions, we will establish the nomenclature here that $E$ is the `\name
entropy.'\footnote{In information theory, `differential entropy' refers to a
distinct quantity \cite{horse}. However, we feel that this information
theoretic application is remote enough from the present context
that our choice of nomenclature here will not lead to any confusion.}

The remainder of the paper is organized as follows:
In section \ref{ads3}, we review the calculation of the Bekenstein-Hawking entropy
of closed curves in the bulk of three-dimensional AdS space in terms of the \name
entropy of a set of intervals in the boundary CFT. However, we provide
a perspective that is distinct from the original presentation in \cite{hole}. In particular,
we introduce the geometric concept of the `outer envelope,' which allows for more
intuitive picture of this construction. In section \ref{subtle}, we also point
out some geometric subtleties, which call for generalizations of both the \name
entropy and the outer envelope.
In section \ref{higher}, we extend these calculations to higher dimensions.
In particular, we study the situation where a time slice in the boundary is covered by a family of
overlapping strips to evaluate the BH entropy \reef{prop0} of a bulk surface. This construction
limits our analysis to cases with planar symmetry, \ie
the profile of the bulk surface can only depend on one of the boundary coordinates. In section \ref{cause},
we consider using causal holographic information as the basis for this construction in higher dimensions but
show that quite generally this approach does not yield finite results. In section \ref{general},
we extend the discussion to more general holographic backgrounds and in section \ref{gene},
we show that these results can be extended to also include bulk gravity theories where
the gravitational entropy has a more general form. In particular, the latter include higher
curvature theories known as Lovelock gravity, as shown in appendix \ref{loveA}.
We close with a brief discussion of our results and future
directions in section \ref{discuss}.

\section{Holographic holes in AdS$_3$ and the outer envelope}
\labell{ads3}

In this section, we discuss some of the key results of \cite{hole}. In particular, the
BH entropy \reef{prop0} for closed curves in three-dimensional
AdS space, \ie $\sbh=($length of curve$)/(4\Gn)$, can be evaluated in terms of the
combination of entanglement entropies given in eq.~\reef{residue}
for the two-dimensional boundary
CFT. However, we will provide a more intuitive geometric description of their construction,
which in particular, makes no reference to accelerated observers in the bulk or time intervals in
the boundary theory. As we describe below,
a key ingredient of our approach will be the
`outer envelope,' which in the simplest cases can be seen as the boundary of the union of
bulk regions associated with each of the boundary intervals \cite{matt}.
However, as we will see in section \ref{subtle}, this simple definition must be generalized
in certain situations.
Another difference from \cite{hole} is that the present calculations will be formulated
in terms of Poincar\'e coordinates, rather than global coordinates.

A central point in the discussion below and in \cite{hole} is a property of entanglement entropy
known as `strong subadditivity' \cite{ssa}, which is an inequality that holds quite generally in
comparing entanglement entropies of various components of a quantum system. In particular,
for two overlapping regions, $I_1$ and $I_2$, in a QFT, this inequality can be expressed as
\beq
S(I_{1}\cup I_2)+S(I_{1}\cap I_2) \le S(I_1)+S(I_2)\,.
\labell{ssa1}
\eeq
Let us recall the holographic proof of this strong subadditivity or rather the proof
that the RT prescription \reef{define} for holographic entanglement entropy satisfies
this inequality \reef{ssa1}.

For simplicity, we assume that the bulk geometry is static (and so may then be easily
analytically continued to a Euclidean spacetime). Now following \cite{head2}, we consider
two overlapping regions, $I_1$ and $I_2$, on a constant (Euclidean) time
slice in the boundary theory. Figure \ref{int1}a illustrates the regions on the boundary
of the AdS spacetime,\footnote{The figure shows a fixed (global) time slice in three-dimensional
AdS, but our discussion of strong subadditivity
applies directly to higher dimensions as well.} as well as the
corresponding extremal surfaces in the bulk which are used
to evaluate the holographic entanglement entropy. In particular, for $S(I_{1})$
and $S(I_2)$, we have the blue arcs, $i_1$ and $i_2$, respectively. Similarly,
$S(I_{1}\cup I_2)$ and $S(I_{1}\cap I_2)$ are evaluated with the RT prescription
\reef{define} using the green arcs, $i_{1\cup2}$ and $i_{1\cap2}$, respectively.
Now the assumption of a static bulk has two simplifying effects. First, as is implicit
in the figure, all of the relevant extremal surfaces lie in the same constant time
slice in the bulk geometry and second, the extremization procedure in eq.~\reef{define}
picks out the bulk surfaces with the minimal surface area, rather than just saddle-points.
As a result of the first property, the two surfaces, $i_1$ and $i_2$, intersect in the bulk
along some codimension-three surface, denoted by the point $p_\textrm{int}$ in figure \ref{int1}a.
Now in this holographic construction, we can consider exchanging the interconnections
of the original surfaces at this intersection and then re-express the right-hand side of
eq.~\reef{ssa1} in terms of the areas of the resulting surfaces, which we denote as
$k_{1\cup2}$ and $k_{1\cap2}$. We illustrate this re-arrangement
with the red ($k_{1\cup2}$) and yellow ($k_{1\cap2}$) arcs in figure \ref{int1}b.
As indicated by the subscripts, $k_{1\cup2}$ and $k_{1\cap2}$ are
homologous to $i_{1\cup2}$ and $i_{1\cap2}$, respectively. However, since the latter are
the extremal surfaces within their respective homology classes, we have  ${\cal A}(i_{1\cup2})<
{\cal A}(k_{1\cup2})$ and ${\cal A}(i_{1\cap2})<
{\cal A}(k_{1\cap2})$, and therefore the desired inequality \reef{ssa1} is satisfied.
Here we might add that a proof of strong subadditivity \reef{ssa1} for holographic
entanglement entropy in nonstatic backgrounds was recently formulated but
is much more elaborate \cite{aron2}.
\begin{figure}[h!]
\centering
\subfloat[]{\includegraphics[width=0.45\textwidth]{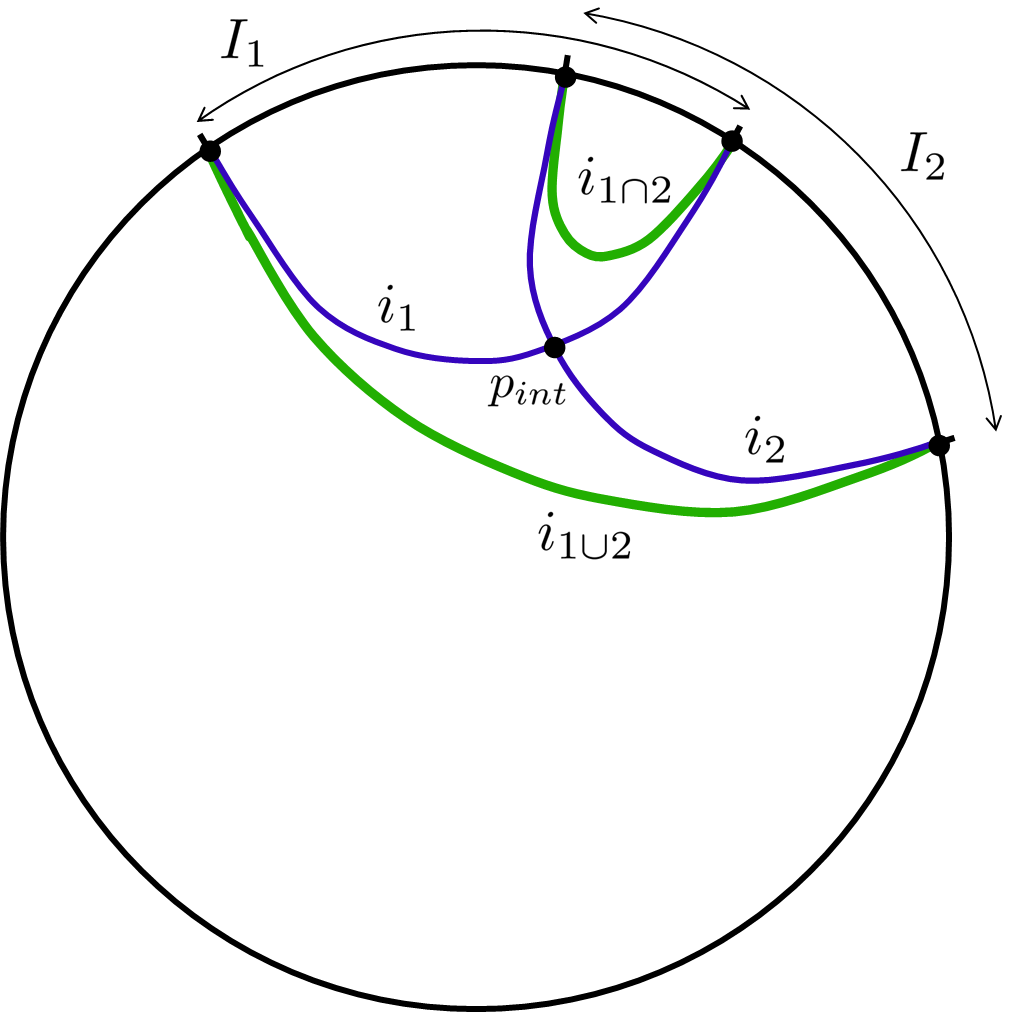}}\qquad
\subfloat[]{\includegraphics[width=0.45\textwidth]{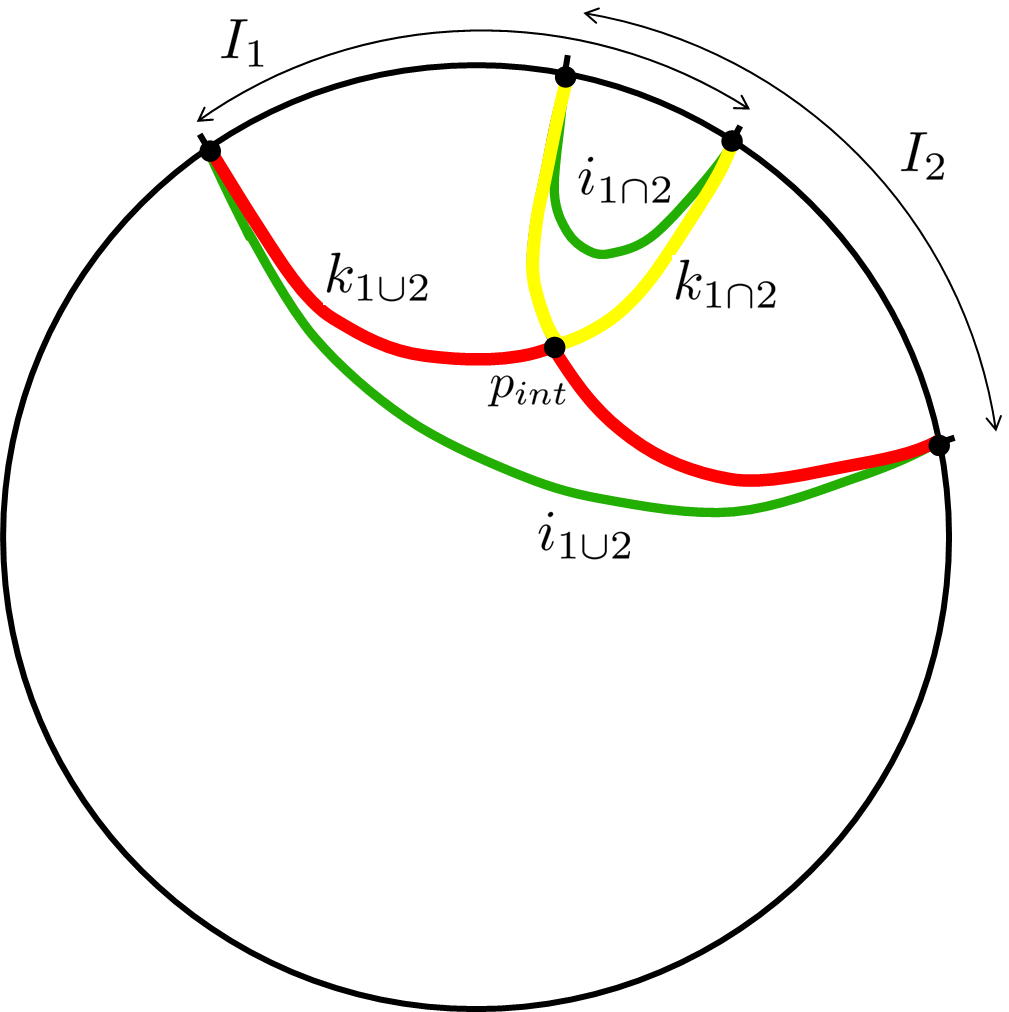}}
\caption{(Color online) Proof of strong subadditivity in a holographic framework:
(a) Two intervals on the boundary of AdS$_3$ in global coordinates.
The blue arcs indicate the geodesics used to evaluate $S(I_{1})$
and $S(I_2)$, while the green arcs are those which determine
$S(I_{1}\cup I_2)$ and $S(I_{1}\cap I_2)$.
(b) Rearranging the interconnection of the blue arcs at their intersection
produces two new curves in the same homology classes as the green
arcs. However, the lengths of the red and yellow curves must be longer than
that of the homologous green arcs.}\labell{int1}
\end{figure}

\begin{figure}[h!]
\centering
\subfloat[]{\label{int2}\includegraphics[width=0.45\textwidth]{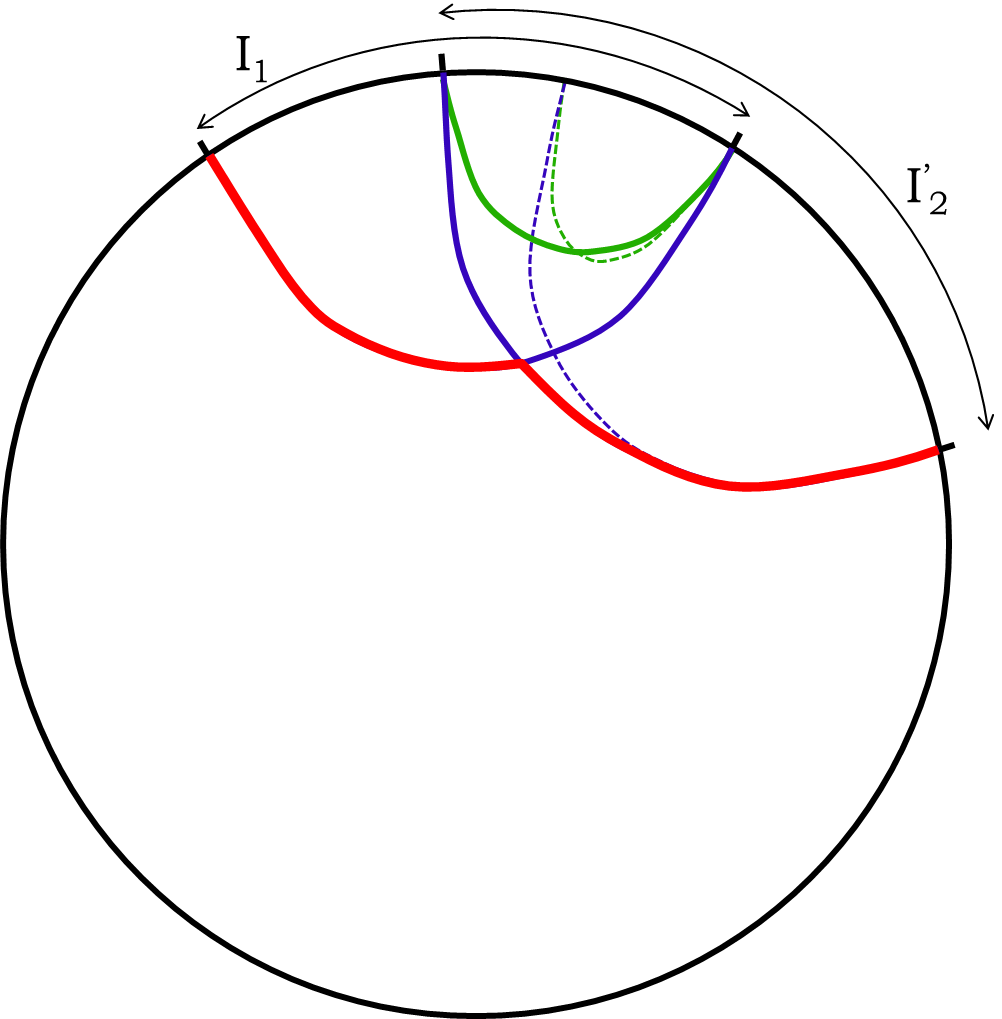}}\qquad
\subfloat[]{\label{int3}\includegraphics[width=0.45\textwidth]{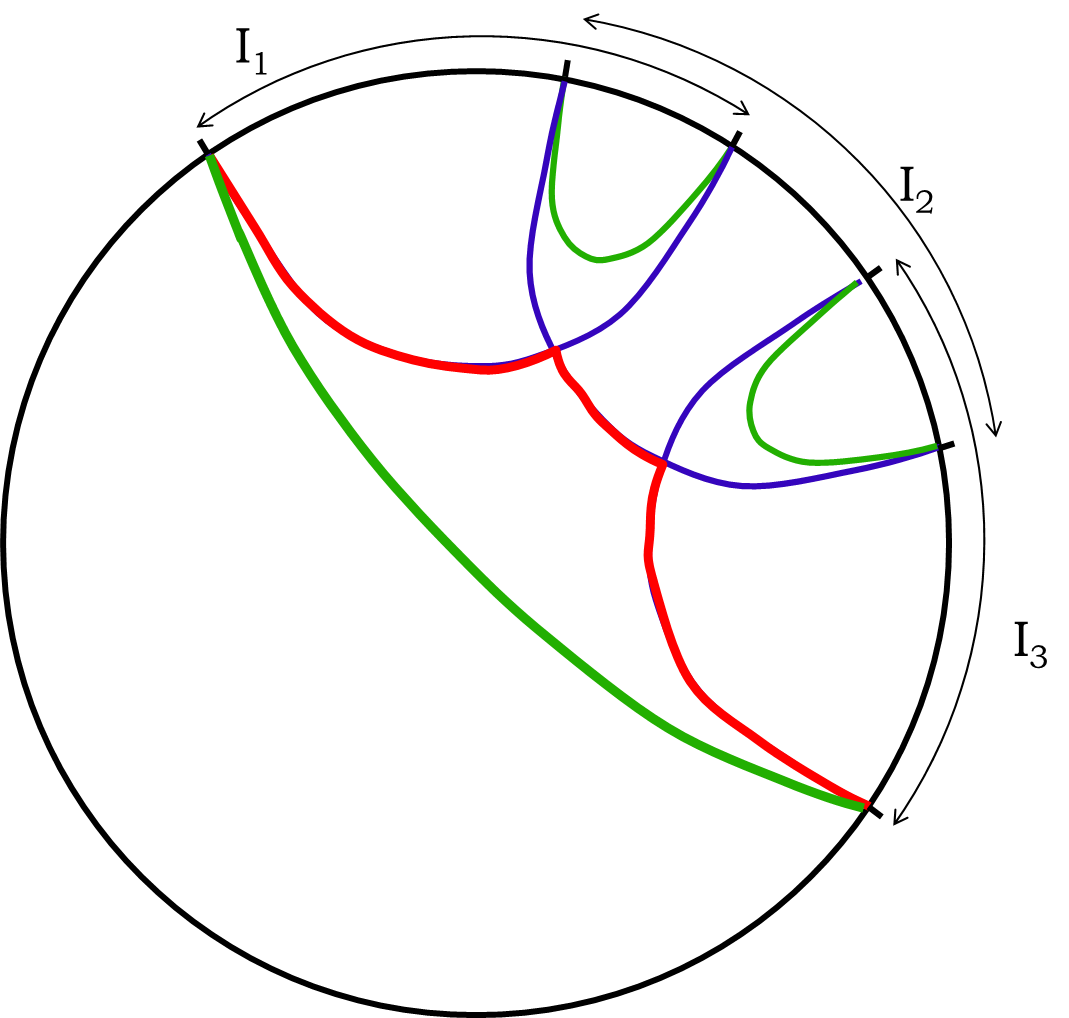}}
\caption{(Color online) (a) This figure illustrates that the outer envelope depends
on the details of the individual boundary intervals, not just their union.
(b) An example with three boundary intervals and the corresponding outer envelope (in red).}
\label{int23}
\end{figure}

For the remainder of the discussion in this section, we will focus on a
three-dimensional bulk spacetime, however, the observations made here are readily
extended to the configurations in higher dimensions that are examined in the subsequent
sections. Returning to figure \ref{int1}b, we will denote the surface $k_{1\cup2}$ as
the `outer envelope.' More generally of a family of intervals $I_k$, we can define the outer
envelope as the boundary of the union of all of the bulk regions enclosed by the geodesics
determining $S(I_k)$ according to the RT prescription \cite{matt}.
Further for our example here, let us denote the Bekenstein-Hawking entropy
\reef{prop0} evaluated on this surface as
 \be
 \hS(I_1,I_2)=\A(k_{1\cup2})/4\Gn\,,
 \labell{post9}
 \ee
which we will loosely refer to as the `entropy of the outer envelope.'
Of course, the endpoints of $k_{1\cup2}$ are defined by the endpoints of the union of
the corresponding boundary intervals. However, as illustrated in figure \ref{int23}a,
the full geometry of the outer envelope is not just a function of $I_1\cup I_2$,
but rather it depends on the details of the partition of this boundary region.
Hence, the entropy $\hS$ is indicated to be a function of $I_1$ and $I_2$ individually
in eq.~\reef{post9}.

Now following the reasoning presented in proving strong subadditivity for holographic
entanglement entropy above, one can easily verify that the following inequalities hold
 \be
S(I_1\cup I_2)\leq\hS(I_1,I_2)\leq S(I_1)+S(I_2)-S(I_1\cap I_2)\,.
\labell{post8}
 \ee
For example, $S(I_1\cup I_2)\leq\hS(I_1,I_2)$ holds because $i_{1\cup2}$ and
$k_{1\cup2}$ are in the same homology class but $i_{1\cup2}$ is the extremal surface
chosen to minimize the BH entropy within this class. Now if one has $n$ consecutive
overlapping intervals on the boundary, these arguments can be extended to establish
the following generalization of eq.~\reef{post8}
 \be
S(\cup I_k)\leq\hS(\lbrace I_k\rbrace)\leq \sum_{k=1}^n S(I_k)-\sum_{k=1}^{n-1}
S(I_k\cap I_{k+1})\,.
\labell{post7}
 \ee
An example of the corresponding surfaces are illustrated in figure \ref{int23}b
for three boundary intervals.

We will primarily be interested in the case where, in fact, the intervals are chosen to
cover the entire boundary, as illustrated in figure \ref{int45}. In this case, we write
the corresponding inequalities as
 \be
S(\cup I_k)\leq\hS(\lbrace I_k\rbrace)\leq \sum_{k=1}^n S(I_k)-\sum_{k=1}^{n}
S(I_k\cap I_{k+1})\,.
\labell{post6}
 \ee
Note that eqs.~\reef{post7} and \reef{post6} are distinct because the second sum in that latter
includes an $n$'th term \cite{hole}, which should be interpreted as $S(I_n\cap I_{n+1})=
S(I_n\cap I_{1})$.  In this scenario where the entire boundary is covered, the outer envelope
forms a closed curve in the bulk. Note that in the case when the bulk geometry is empty AdS
space, as illustrated in figure \ref{int45}a, the entanglement entropy for the union of all
the intervals vanishes. This result arises from the
bulk perspective since the prescription for holographic entanglement entropy \reef{define}
instructs us to find an extremal surface which is homologous to the entire boundary.
Hence we are considering closed surfaces in the bulk but upon extremizing within this
class, the minimal area is found when the surface simply shrinks to a point and the area vanishes.
In contrast, no extremization appears in the construction of the outer envelope and so
the corresponding entropy  $\hS(\lbrace I_k\rbrace)$ remains finite.
From the boundary perspective, the previous vanishing is natural because empty AdS$_3$ is dual to
the vacuum of the boundary CFT and so the corresponding entropy vanishes.
In fact, when the bulk is dual to any pure state
in the boundary theory, the entanglement entropy for the union of all the intervals
must similarly vanish, \ie $S(\cup I_k)=S(|\psi\rangle\langle\psi|)=0$. Of course,
as illustrated in figure \ref{int45}b, if the bulk is a (stationary) black hole geometry, then
$S(\cup I_k)$ is non-vanishing. Here, the
desired extremal surface corresponds to (the bifurcation surface of) the horizon and the
corresponding entanglement entropy is just the thermodynamic entropy of the dual thermal
ensemble in the boundary theory. In this instance, the outer envelope still defines a larger
entropy, \ie $\hS(\lbrace I_k\rbrace)\geq S(\cup I_k)=S(\rho_\textrm{therm})$.
\begin{figure}[h!]
\centering
\subfloat[]{\label{int4}\includegraphics[width=0.45\textwidth]{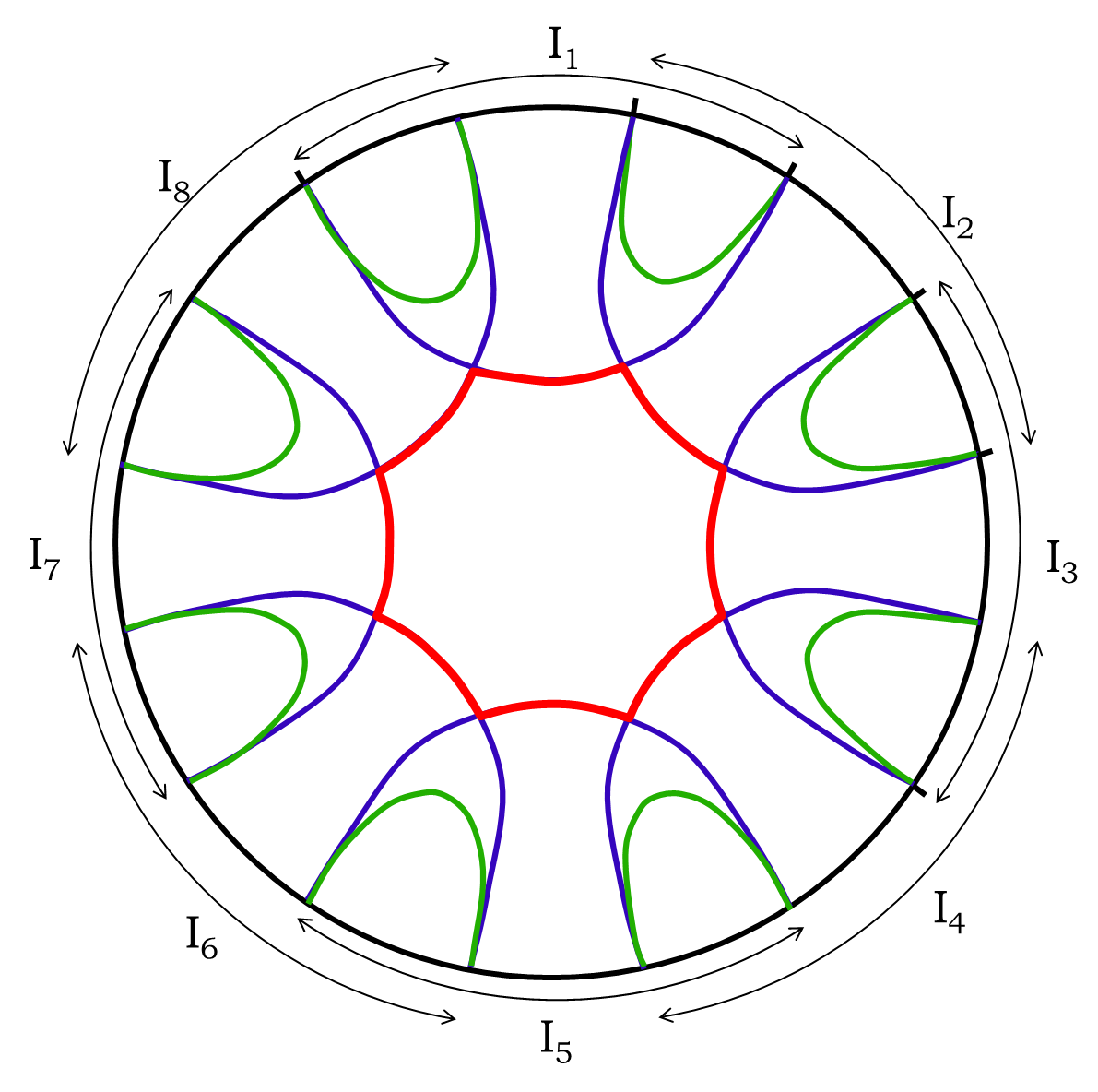}}\qquad
\subfloat[]{\label{int5}\includegraphics[width=0.45\textwidth]{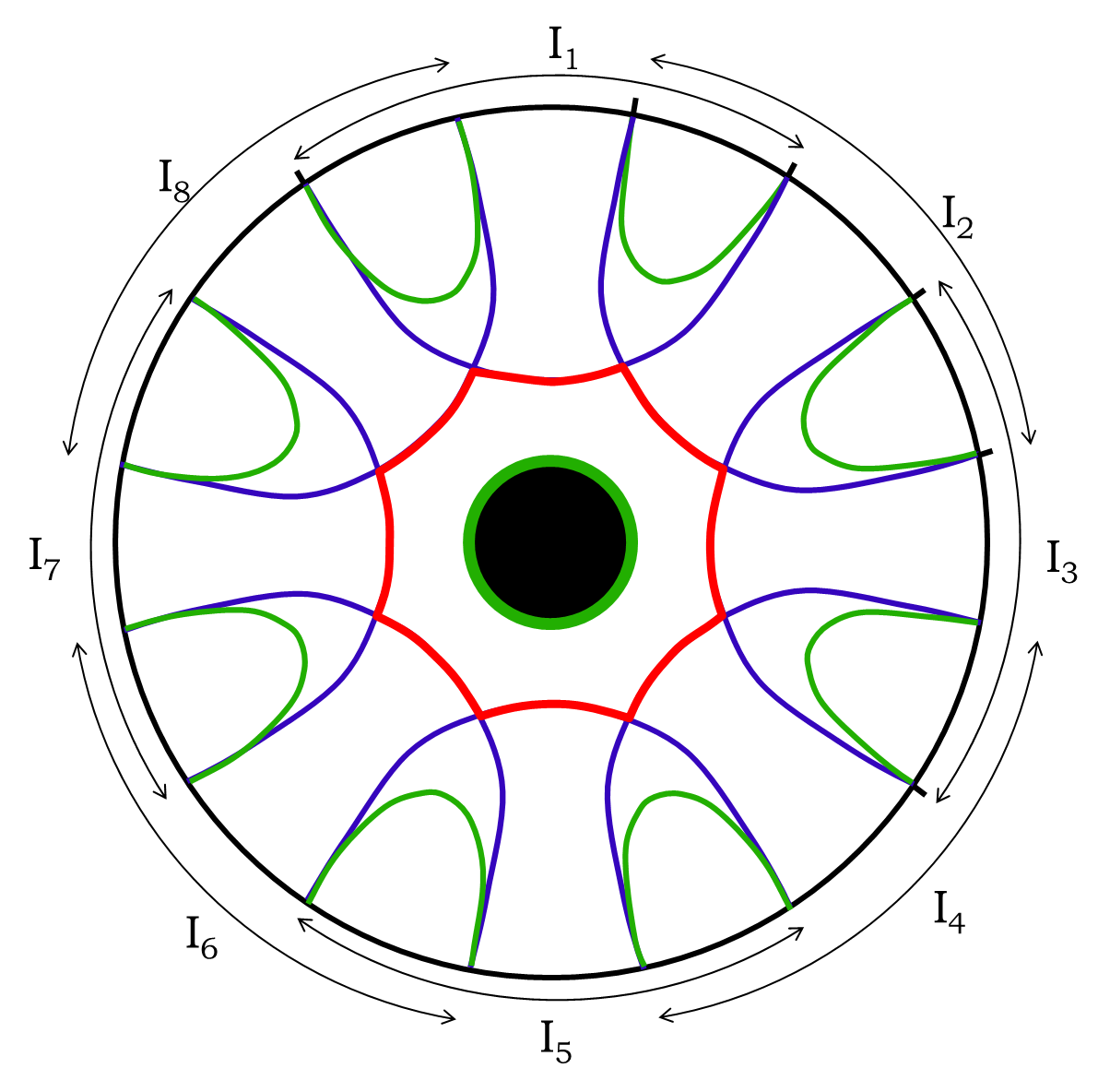}}
\caption{(Color online) Eight intervals and their outer envelope, which forms a closed curve in the bulk,
(a) in empty AdS space and (b) in an AdS black hole spacetime. In case (a), the entanglement
entropy of the global boundary state vanishes while it is non-vanishing in case (b). }
\labell{int45}
\end{figure}

\begin{figure}[h!]
\centering
\subfloat[]{\includegraphics[width=0.45\textwidth]{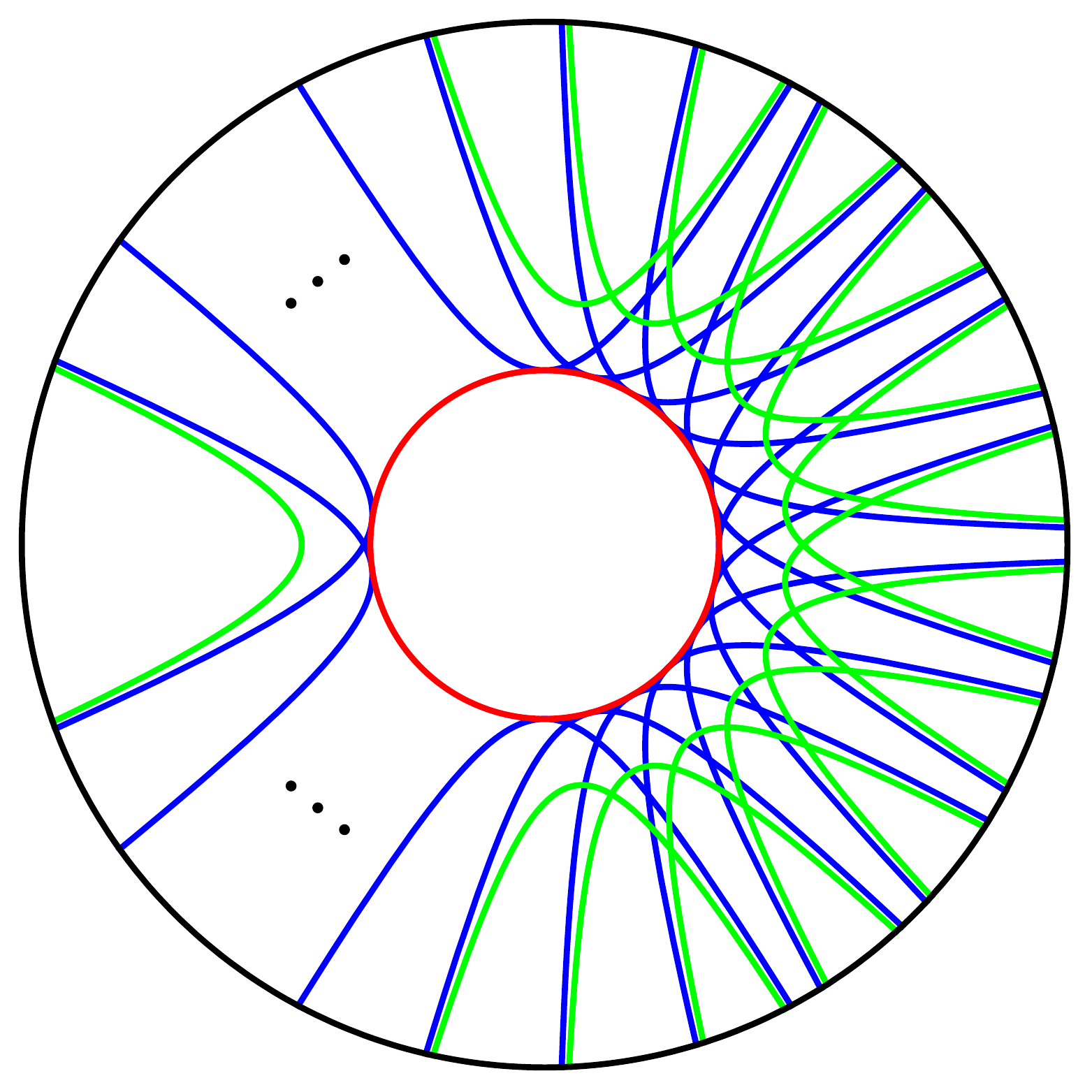}}\qquad
\subfloat[]{\includegraphics[width=0.45\textwidth]{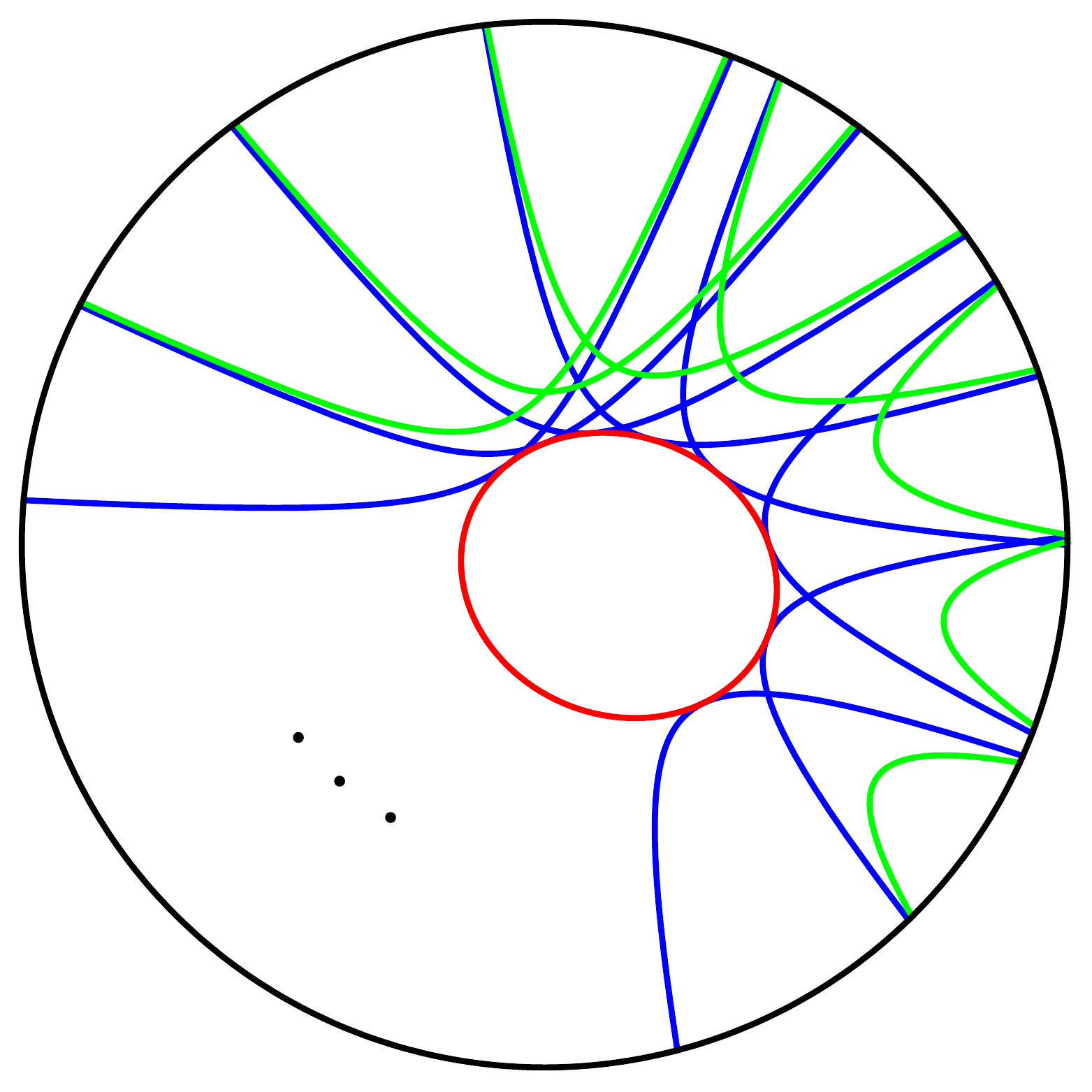}}
\caption{(Color online) (a) In the continuum limit of many identical intervals,
the outer envelope becomes a circle of a fixed radius.
(b) The continuum limit of many intervals whose length varies continuously
produces a smooth outer envelope with a profile that varies in the bulk.}
\labell{hole}
\end{figure}
Above, we have introduced a class of closed curves in the bulk which are constructed
as the outer envelope of a series of extremal surfaces determining the holographic
entanglement entropies of
some ordered set of intervals which partitions an entire time slice of the boundary.
In general, eq.~\reef{post6} indicates that the BH entropy of
these closed curves is bounded below by the entanglement entropy of the boundary state
and bounded above by a certain combination of entanglement entropies of the boundary
intervals and their intersections. Now following \cite{hole}, we extend these observations
with the following construction (which be described in detail below): First, we keep the length
of the individual intervals fixed but take the number of intervals equally spaced
around the boundary to infinity. In this limit,
the outer envelope becomes a smooth circle in the bulk with a fixed radius $R$,
as shown in figure \ref{hole}a, and
the corresponding Bekenstein-Hawking entropy is simply $\hS(\lbrace I_k\rbrace)=
2\pi R/4\Gn$. The remarkable discovery in \cite{hole} is that the second
inequality in eq.~\reef{post6} is in fact saturated in this limit, namely
 \be
\frac{2\pi R}{4\Gn}=\sum_{k=1}^\infty \[\,S(I_k)-S(I_k\cap I_{k+1})\,\]
\,,\labell{surprise1}
 \ee
Further, with an appropriate extension of this continuum limit illustrated
in figure \ref{hole}, one finds the same
equality holds for a general closed curve in the bulk,
 \be
\frac{\A(\textrm{bulk curve})}{4\Gn}
=\sum_{k=1}^\infty \[\,S(I_k)-S(I_k\cap I_{k+1})\,\]\,.
 \ee
Hence the two-dimensional boundary theory appears to have `observables' corresponding to
the BH entropy of arbitrary closed curves in the bulk of AdS$_3$. Before
proceeding to higher dimensions, let us describe the construction for AdS$_3$ in more detail,
 for the case where it is adapted to Poincar\'e coordinates.

\subsection{Holographic holes in AdS$_3$}
\labell{hole3}

To begin, recall the AdS$_3$ written in Poincar\'e coordinates
 \be
ds^2=\frac{L^2}{z^2}(dz^2-dt^2+dx^2)\,.
\labell{metric3}
 \ee
Now if we wish to evaluate the holographic entanglement entropy of an
interval of width $\Delta x$, the extremal surface simply takes the form
of a semi-circle in these coordinates \cite{rt1,rt2}, \ie
 \be
z^2 + x^2 = (\Delta x/2)^2\,,
\labell{circle}
 \ee
and evaluating the length of this extremal curve yields
 \be
S(\Delta x)=\frac{L}{2\Gn}\ \log\!\(\frac{\Delta x}{\delta}\)\,,
\labell{cc3}
 \ee
where $\delta$ is the short-distance cut-off in the boundary
theory, which is introduced with a cut-off surface in the bulk at $z=z_\textrm{min}=\delta$.
Of course, upon substituting $c=3L/2\Gn$, this holographic result \reef{cc3} reproduces the
universal result which applies for any two-dimensional CFT \cite{frank1,cardy1}.

\begin{figure}[h!]
\begin{center}
\includegraphics[width=0.7\textwidth]{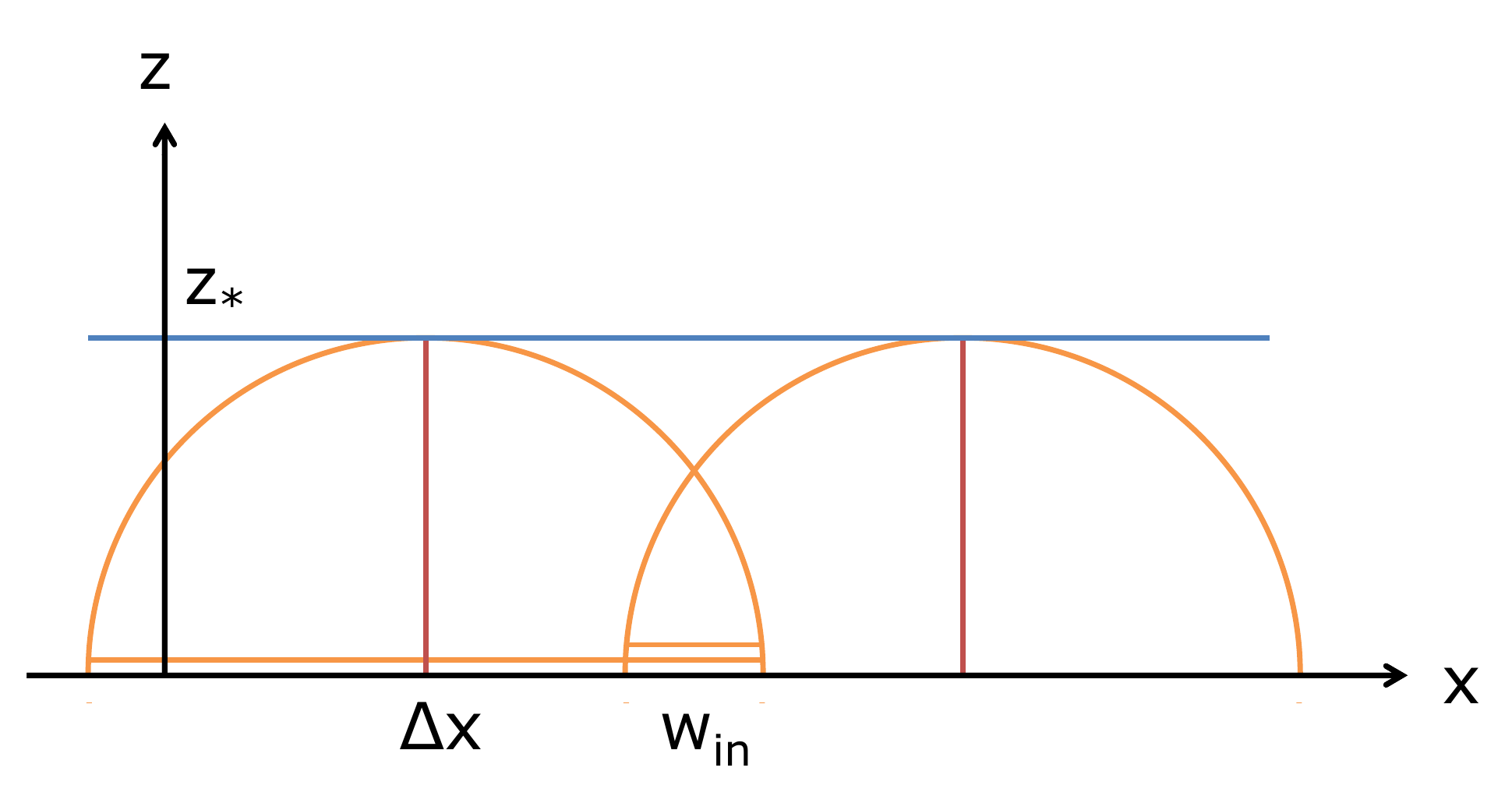}
\caption{(Color online) A bulk surface with a constant profile $z=z_*$. The two intersecting semi-circles
of radius $r=z_*$ in AdS$_3$ are the extremal bulk surfaces determining the holographic
entanglement entropy for two overlapping boundary intervals of length
$\Delta x=2z_*$.}
\label{fig3}
\end{center}
\end{figure}

For simplicity, let us begin by considering the Bekenstein-Hawking entropy for
a surface in the bulk at a fixed $z=z_*$, as illustrated in
figure \ref{fig3}.  To regulate the area of this surface, we will impose that the $x$
direction is periodic with period $\ell_1$. One should think of the latter as some
infrared regulator scale and so we assume that $\ell_1\gg\Delta x$.
Now given the bulk metric \reef{metric3}, we find
the BH entropy of the surface is given by
 \be
 \frac{\A(z=z_*)}{4\,\Gn}=\frac{L\ \ell_1}{4\Gn\,z_*}\,.
 \labell{goal0}
 \ee

Now to begin, we consider a series of $n$ equally spaced intervals $I_k$  with a fixed width
$\Delta x$, which cover the boundary. We choose $\Delta x=2 z_*$
so that the corresponding extremal semi-circles \reef{circle} in the bulk
are all tangent to the desired surface at $z=z_*$. Now, the intuition is that the
latter surface emerges as the outer envelope of these semi-circles in the `continuum'
limit where $n\to\infty$. Hence, we first confirm that the entropy formula \reef{goal0} is
reproduced by $\hS(\lbrace I_k\rbrace)$ in the continuum limit. As illustrated in figure \ref{fig3},
it is useful to chose an angular coordinate along the semi-circles with:
 \beq
z=z_*\,\cos\theta\,,\qquad x=x_{c,k}+z_*\,\sin\theta\,,
 \labell{angle}
\eeq
where $x_{c,k}$ is the midpoint of the corresponding interval $I_k$ on the boundary.
With $n$ intervals on the boundary,
the spacing between, \eg their midpoints is simply given by $\ell_1/n$ and hence
to determine the contribution of an individual semi-circle to the full length of the outer
envelope, we must integrate $\theta$ over the range $[-\theta_0,\theta_0]$ where
 \be
\sin\theta_0=\frac{\ell_1}{2n\,z_*}\,.
 \labell{angle0}
 \eeq
Then the Bekenstein-Hawking entropy associated with the full outer envelope becomes
 \beqa
\hS(\lbrace I_k\rbrace) &=& \frac{n}{4\Gn}\,2\,\int_{0}^{\theta_0} d\theta
\frac{L}{\cos\theta}
\nonumber\\
&=&\frac{n L}{4\Gn}\ \log\!\(\frac{1+\frac{\ell_1}{2n\,z_*}}{1-\frac{\ell_1}{2n\,z_*}}
 \)\,.
 \labell{fulls}
 \eeqa
Finally, it is straightforward to see that in the limit $n\to\infty$, this result \reef{fulls}
simplifies to precisely the desired entropy given in eq.~\reef{goal0}.

Now we turn to the \name entropy \reef{residue} of the same family of intervals $I_k$.
Recall the width of each interval was $\Delta x=2z_*$. Hence with $n$ equally spaced
intervals on the boundary of length $\ell_1$, we find that
the length of the intersections $I_k\cap I_{k+1}$ is given by
 \be
w_\textrm{in}=2z_*-\frac{\ell_1}{n}\,,
 \labell{win}
 \ee
as is seen in figure \ref{fig3}. Hence combining the above results, the \name
entropy \reef{residue} becomes
 \beqa
E(\lbrace I_k\rbrace) =\sum_{k=1}^n \[\, S(I_k)-S(I_k\cap I_{k+1})\,\]&=&
 \frac{n L}{2\Gn}\ \[\log\!\(\frac{2z_*}{\delta}\)
-\log\!\(\frac{w_\textrm{in}}{\delta}\)\]
\nonumber\\
&=&-\frac{n L}{2\Gn}\ \log\!\(1-\frac{\ell_1}{2nz_*}\)\,.
\labell{course0}
 \eeqa
Again, we can easily show that in the limit $n\to\infty$, this result \reef{course0}
simplifies to the desired entropy in eq.~\reef{goal0}.

As an aside, let us combine eqs.~\reef{fulls} and \reef{course0} to establish
 \beq
E(\lbrace I_k\rbrace)- \hS(\lbrace I_k\rbrace)
=-\frac{n L}{4\Gn}\ \log\!\[1-\(\frac{\ell_1}{2nz_*}\)^2\]\ge 0\,.
\labell{finiten}
 \eeq
Hence eq.~\reef{finiten} explicitly shows that at finite $n$,
$E(\lbrace I_k\rbrace)> \hS(\lbrace I_k\rbrace)$, as expected, and it is only in the
continuum limit with $n\to\infty$, that we have $E(\lbrace I_k\rbrace)= \hS(\lbrace I_k\rbrace)$.

\begin{figure}[h!]
\begin{center}
\includegraphics[width=0.7\textwidth]{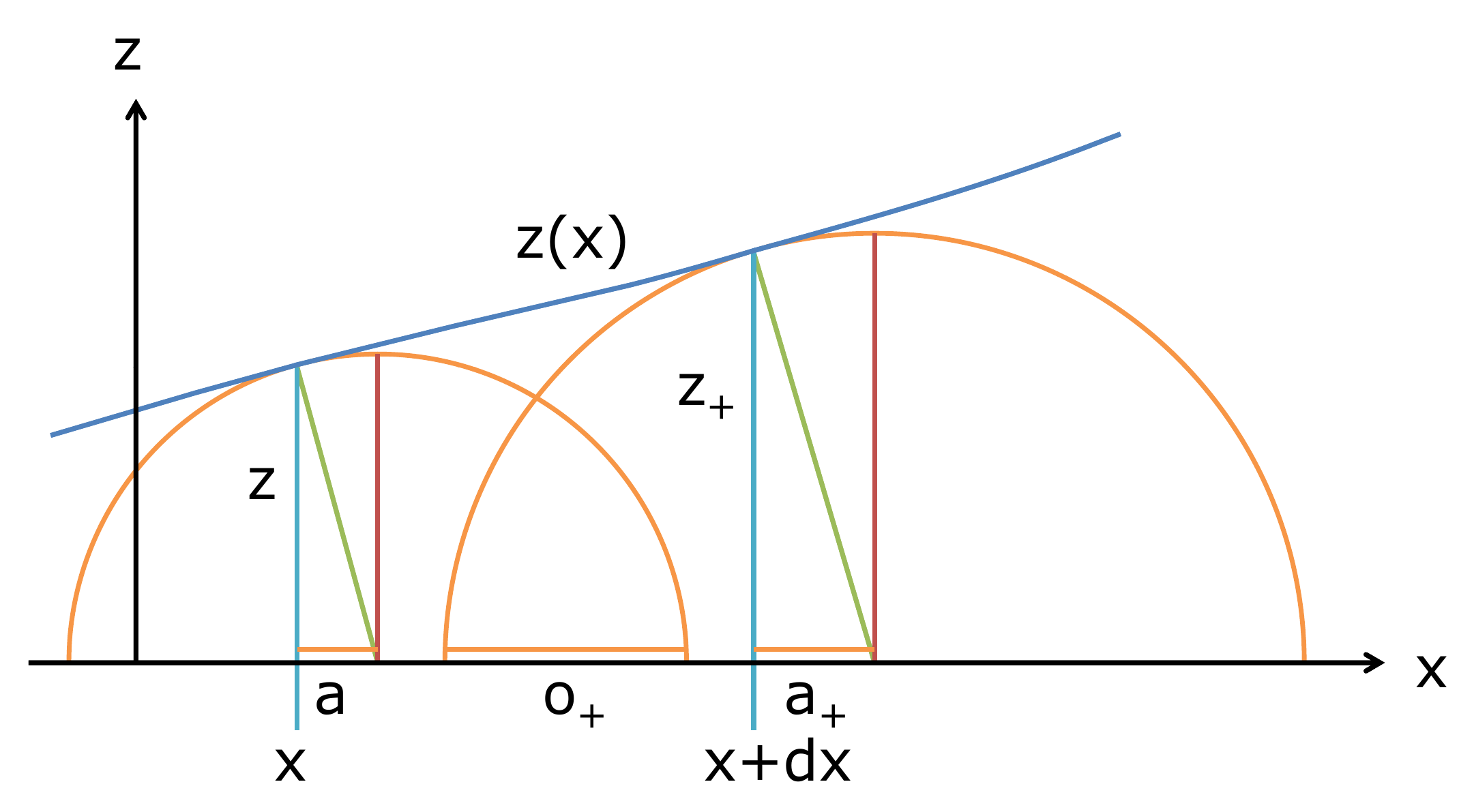}
\caption{(Color online) A bulk surface with a varying profile $z=z(x)$. The radius of the semi-circles
tangent to this surface varies with $x$ and hence the length of the corresponding boundary
intervals is also a function of $x$. Further, the center of the boundary interval is displaced
from the tangent point along the $x$-axis by some distance $a(x)$.}
\label{fig4}
\end{center}
\end{figure}

Next let us consider a bulk surface of varying profile $z(x)$, as illustrated
in figure \ref{fig4}. In this case,
evaluating the Bekenstein-Hawking formula \reef{prop0} on this curve yields
 \be
 \frac{\A(z=z(x))}{4\,\Gn}=\frac{L}{4\Gn}\
\int_0^{\ell_1}\!dx\,\frac{\sqrt{1+z'^2}}{z}\,,
 \labell{goal1}
\ee
where again we have assumed that the $x$ direction is periodic with period $\ell_1$.

In this case, we will work directly in the continuum limit with an infinite number
of boundary intervals. The key idea is to find a family of intervals $I_k$ such that
there is a dual semi-circle in the bulk which is tangent to each point along the chosen surface.
The geometry for two neighboring intervals is shown in figure \ref{fig4}.
Here we choose points on the bulk surface which are separated infinitesimally
along the $x$ direction, \ie the points at $x$ and $x\pm dx$ which sit at $z=z(x)$
and $z_\pm =z(x\pm dx)$ in the bulk. Note that generally the midpoint of the corresponding intervals
is displaced from the position of the tanget point because of the nonvanishing slope of
$z(x)$. We denote this shift as $a(x)$, as shown in figure \ref{fig4}, and examining
the geometry, we find
 \beq
 z'(x)=\frac{a(x)}{z(x)}\,.
 \labell{slope}
 \eeq
Now let us denote the width of the corresponding intervals as $\Delta x(x)$ and since $\Delta x(x)
=2\,r(x)$ where $r(x)$ is the radius of the corresponding semi-circle in the bulk, we find
 \be
\Delta x(x)=2\,z(x)\,\sqrt{1+z'(x)^2}\,,
 \labell{radius}
 \ee
using eq.~\reef{slope}. Finally, if we use $o_\pm$ to denote the overlaps between the
interval corresponding to the point at $x$ and those at $x\pm dx$, then we can show
 \be
o_\pm=\frac{1}{2}\(\Delta x(x)+\Delta x(x\pm dx)\)\pm\(a(x)-a(x\pm dx)\)
-dx\,.
 \labell{over2}
 \ee
Writing eq.~\reef{slope} as $a=z\,z'$, we can expand the combination of shifts appearing
above to first order in $dx$ to find
 \be
\pm\(a(x)-a(x\pm dx)\)=-a'dx=-(z'^2+z\,z'')dx\,.
 \labell{face3}
 \ee
Combining this expression with eq.~\reef{radius}, the overlaps in eq.~\reef{over2} can be
written as
 \be
o_\pm=2z\sqrt{1+z'^2}-dx\(1+z'^2+zz''\)\pm\frac{1}{2}\Delta x'dx\,.
 \labell{over3}
 \ee
Note that as may have been expected to leading order in the continuum limit, the overlap
between neighboring intervals is complete, \ie $o_\pm=\Delta x(x)+O(dx)$.

In the above analysis (and in figure \ref{fig4}), we implicitly assumed that the slope
of the profile was positive (\ie $z'\ge0$) at the points $x$ and $x\pm dx$. However, our
results are readily adapted to also cover the case of negative slopes. In particular, when $z'$
is negative, we see that $a$ changes its sign from eq.~\reef{slope}. But the $\pm$ in front of
$(a-a_\pm)$ in eq.~\reef{over2} also needs to be flipped. Hence the final formula \reef{over3}
covers the case of negative slopes as well.\footnote{An analogous result will also apply in our analysis
for higher dimensions in the subsequent sections.}
Henceforth for simplicity, we will assume  that $z'$ is positive without loss of generality.

Now following \cite{hole}, the final steps in the analysis is simplified if we replace the
\name entropy \reef{residue} with the following `averaged' expression
 \be
E=\sum_{k=1}^n \[\, S(I_k)-\frac12 S(I_k\cap I_{k+1})-\frac12 S(I_{k-1}\cap I_{k})\,\]
\,. \labell{residue3}
 \ee
Then combining eq.~\reef{over3} with the expression for the holographic entanglement entropy
\reef{cc3}, we find
 \beqa
S(I_k)-\frac12 S(I_k\cap I_{k+1})&-&\frac12 S(I_{k-1}\cap I_{k})=\frac{L}{4\Gn}\ \log\(
\frac{\Delta x^2}{o_+\,o_-}\)
\nonumber\\
&\simeq&\frac{L}{4\Gn}\frac{1+z'^2+zz''}{z\,\sqrt{1+z'^2}}\,dx +O(dx^2)\,.
 \labell{integrand}
 \eeqa
Hence the \name entropy \reef{residue3} becomes
 \beqa
E(\lbrace I_k\rbrace)&=&\frac{L}{4\Gn}\int_0^{\ell_1}
\frac{1+z'^2+zz''}{z\,\sqrt{1+z'^2}}\,dx
\nonumber\\
&=&\frac{L}{4\Gn}\int_0^{\ell_1}
\[\frac{\sqrt{1+z'^2}}{z}+\frac{z''}{\sqrt{1+z'^2}}\]dx
\nonumber\\
&=&\frac{L}{4\Gn}\[\int_0^{\ell_1}\! dx\,\frac{\sqrt{1+z'^2}}{z}+
{\rm arcsinh}(z')\biggl|_0^{\ell_1}\]\,. \labell{9}
 \eeqa
Of course, the final contribution cancels due to the periodic boundary conditions which
were chosen at the outset of our calculation. However, one can see that this term will
vanish with other choices of boundary conditions, as well. For instance, if the $x$ direction was of infinite extent, it
would suffice to impose $z'\to 0$ as $x\to\pm\infty$. In any event, the remaining integral
precisely matches the expression in eq.~\reef{goal1} for the BH entropy \reef{prop0}
of the bulk surface with profile $z(x)$.

\subsection{Some geometric subtleties}
\labell{subtle}

Our previous discussion makes the implicit assumption that the profile $z(x)$
of the bulk curve varies relatively slowly. In particular, we assume that at each
point along the curve, the curvature of the profile is small enough that the curve
remains outside of the corresponding semi-circle that is tangent at this point.\footnote{In
the context of the AdS geometry, we could express this constraint in terms of the proper
acceleration of the bulk curve. The semi-circles are extremal and so correspond to spatial
geodesics in the bulk geometry. Therefore the proper acceleration vanishes for all of these curves.
Hence we need only demand that the profile $z(x)$ corresponds to a bulk curve with
positive (or vanishing) proper acceleration.} Therefore we will examine next
the changes in the previous construction when this assumption no longer holds.
One observation is that we will have to revise the definition of the `outer envelope'
introduced at the beginning of this section to extend our discussion to cover
these situations. However, we will first examine a simple example to gain
some qualitative understanding of the (unexpected) behavior that arises in this situation.

\begin{figure}[h!]
\begin{center}
\includegraphics[width=0.9\textwidth]{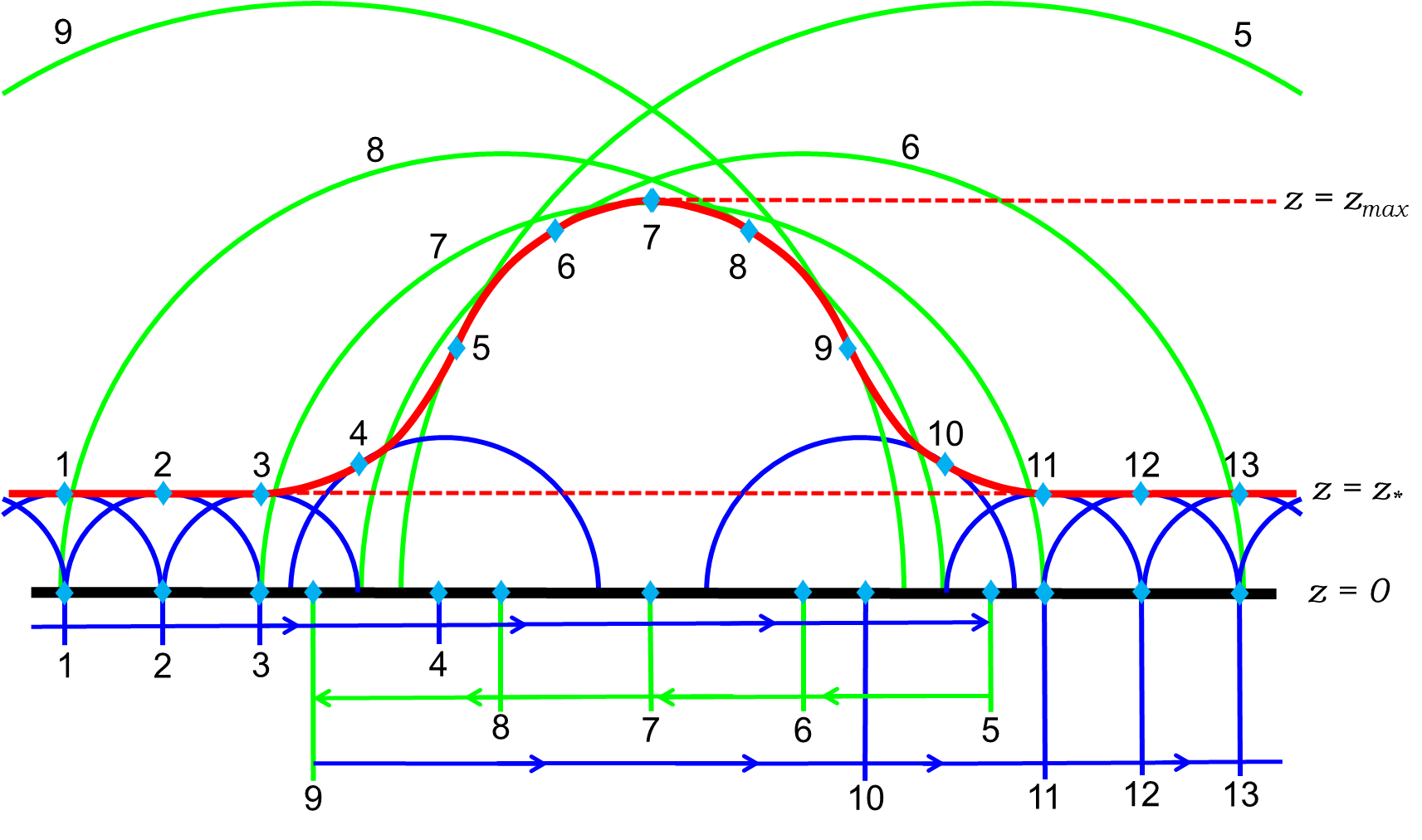}
\caption{(Color online) A bulk surface (red) which is flat with
$z =z_*$ apart from a bump rising to $z=z_\textrm{max}$ and then returning to
$z=z_*$ over a fairly narrow interval. The semi-circles tangent to points
regularly spaced in $x$ along this curve are shown in blue and green. The
green semi-circles extend into the bulk beyond the surface. The centers
of the corresponding boundary intervals are also shown.}
\label{bump}
\end{center}
\end{figure}

In figure \ref{bump}, we illustrated a simple bulk curve where some of the
semi-circles tangent to this surface extend further into the bulk beyond the curve. The (red) profile shown there is flat with
$z =z_*$ apart from a bump, where it rises to $z=z_\textrm{max}$ and then returns to
$z=z_*$ over a fairly narrow interval. The figure also shows the semi-circles that are
tangent to points on the profile that are regularly spaced along the $x$-axis. The (green) semi-circles
tangent to the points near the peak of the bump clearly extend into the bulk beyond the
(red) curve and so these are the points of primary interest here. The endpoints for the corresponding intervals on the
AdS boundary at $z=0$ (which corresponds to the thick black line in the figure) are defined by
the intersection of the semi-circles with $z=0$. The center of each
semi-circle, which is also the center of the corresponding boundary interval, is also indicated in
figure \ref{bump} for each of the points along the bulk curve. In regions where the bulk profile is flat, the
$x$ positions of the tangent point in the bulk and the center of the corresponding interval on the
boundary actually coincide. However, when the profile starts to curve up into the bulk, we can see
that the center points on the boundary begin to `accelerate'
ahead of the tangent points. This acceleration stops where the slope $dz/dx$ reaches
its maximum, \ie at the point denoted 5. This point is also the first one for which the tangent semi-circle
is not contained entirely within the bulk profile. As indicated in the figure, at this stage, the center
points actually begin to move backwards along the $x$-axis --- even though, the corresponding tangent points
in the bulk are still moving forward. This reverse motion of the center points continues for the tangent
points across the peak of the bump, which corresponds to all of those points for which the tangent
semi-circle extends beyond the bulk profile. This process ends where the slope $dz/dx$ is most
negative, \ie at the point marked 9. At this tangent point, the corresponding center point on the
boundary is behind along the $x$-axis. For the subsequent points, the tangent semi-circles are all
contained within the bulk curve and the center points are again moving forward towards larger values of $x$.

The example in figure \ref{bump} seems to connect the tangent semi-circle extending
beyond the bulk surface and the backward motion of the center point of the boundary interval.
So we would like use our results in the previous subsection to verify that this
behavior above is, in fact, a general property of this construction.
First, if we consider the point on the bulk curve to be at $x$, then the center of
the corresponding boundary interval is given by
 \be
x_c(x)=x+a(x)=x+z\,z'\,,
 \labell{center}
 \ee
using eq.~\reef{slope}. Given this expression, we see that $x_c=x$ if and only if
$z'=0$ while $z'>0$ yields $x_c>x$ and $z'<0$ yields $x_c<x$. We can also
evaluate the derivative
 \be
x_c' = 1+z'^2 + z\,z''\,,
 \labell{movec}
 \ee
and next we would like to show that $x_c'<0$ implies that the corresponding semi-circle
in the bulk extends beyond the curve $z(x)$. A Taylor expansion of the bulk profile around
some value of $x$ yields
 \be
z(x+\delta x)= z(x) + z'(x)\,\delta x+\frac{1}{2}z''(x)\,\delta x^2+\cdots\,.
 \labell{expand0}
 \ee
Now the semi-circle, which is tangent to the curve at $x$, is described by the following
profile
 \be
z_{sc}(\tilde x; x)=\sqrt{r(x)^2-(\tilde x-x_c(x))^2}\,,
 \labell{semi8}
 \ee
where the radius is given by $r(x)^2 = z(x)^2+a(x)^2$. Hence using our previous results,
we find an expansion about $x$ yields, \ie with $\tilde x=x+\delta x$,
 \be
z_{sc}(x+\delta x;x)= z(x) + z'(x)\,\delta x-\frac{1}{2}\left( \frac{1+z'^2}{z} \right)\,\delta x^2+\cdots\,.
 \labell{expand1}
 \ee
Comparing eqs.~\reef{expand0} and \reef{expand1}, we see that the first two terms in these expansions
match, which of course was ensured by the construction. Hence, the question of whether or not the semi-circle
remains below the bulk surface is determined, at least locally, by the quadratic term in the expansions.
In particular, the semi-circle extends beyond the curve if
 \be
z''<-\frac{1+z'^2}{z}\,.
 \labell{extend0}
 \ee
Now comparing eqs.~\reef{movec} and \reef{extend0}, we see that the above inequality corresponds
precisely to the condition for $x_c'<0$, as expected.

\begin{figure}[h!]
\begin{center}
\includegraphics[width=0.5\textwidth]{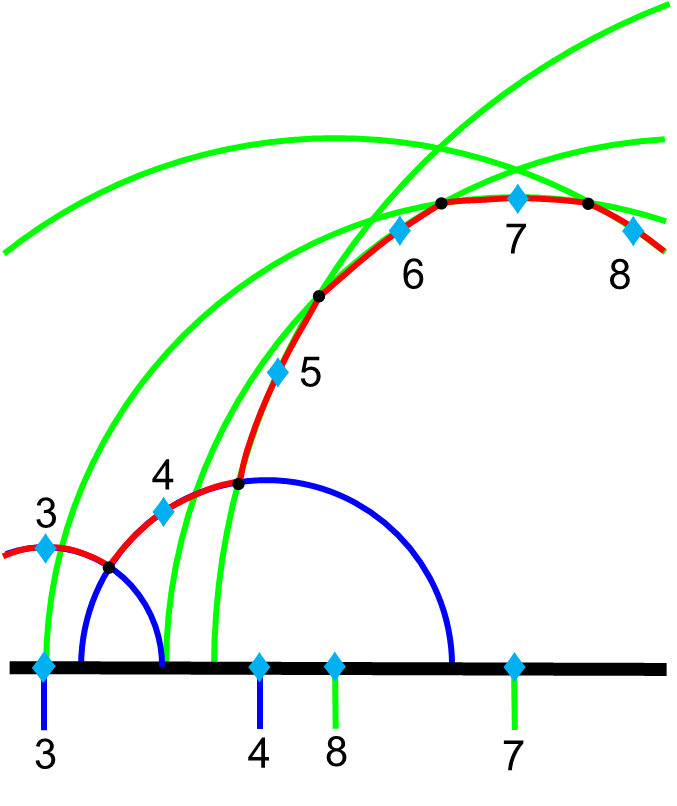}
\caption{(Color online) A portion of figure \ref{bump}, which illustrates that segments
of the tangent semi-circles can still be used to give a good approximation to the desired
bulk curve. Hence a suitable definition for the `outer envelope' is simply the union of these
segments.}
\label{fig10}
\end{center}
\end{figure}
We originally defined the `outer envelope,' as the curve in the bulk is
comprised of the outermost segments of the
extremal surfaces defining the entanglement entropies appearing in the \name entropy \reef{residue}.
While this definition works in the simplest cases, it does not apply in the present case where these
extremal surfaces extend beyond the bulk curve. However, as shown in figure \ref{fig10}, the bulk
curve is still well approximated piece-wise by various segments of these extremal curves. If we examine
this figure, we see the appropriate segments are simply those which extend along the extremal curves
from the tangent point for a given interval $I_k$ to the first intersection with the extremal curves
for the adjacent intervals, \ie $I_{k-1}$ and $I_{k+1}$. Of course, these are perhaps the collection of segments
which intuitively would give a good approximation to the bulk curve. Note that we will still refer to
this collection as the `outer envelope,' even though this nomenclature does not always give
an accurate description of the union of these segments.

\begin{figure}[h!]
\begin{center}
\includegraphics[width=0.7\textwidth]{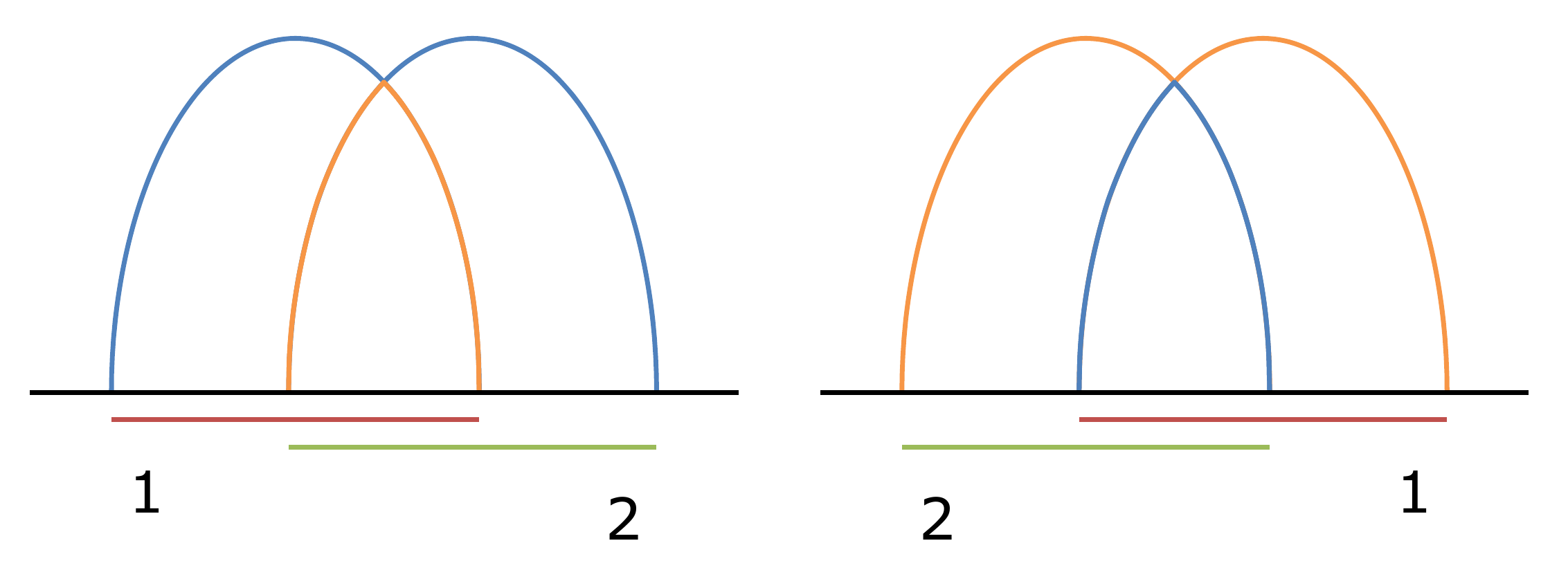}
\caption{(Color online) Generalized outer envelope of two intervals indicated by the blue curve.}
\label{fig9}
\end{center}
\end{figure}

This case where the semi-circles extend beyond the bulk curve also calls for us to revise
our definition of the \name entropy. Recall that this situation also corresponds to the center of
the corresponding intervals moving in the negative direction along the $x$-axis. Hence let us
consider the simplest example of two overlapping intervals, $I_k$ and $I_{k+1}$,
for which the corresponding extremal curves in the bulk intersect, as shown in figure \ref{fig9}.
Let introduce the notation, $x_{L,k}$ and $x_{R,k}$, to denote the left and right end-points of
the interval $I_k$. The `standard' case with $x_c'>0$ is shown on the left
of the figure, while the case with $x_c'<0$ is illustrated on the right.  Note that
in both cases, the outer envelope is comprised of the two segments of the bulk
semi-circles which extend from $x_{L,k}$ to $x_{R,k+1}$. Motivated by discussion of strong
subadditivity, in both cases, we would bound the corresponding BH entropy
$\hS$ by taking the sum of the entanglement entropies $S(I_k)$ and $S(I_{k+1})$ and
subtracting of the entanglement entropy for the interval extending from  $x_{R,k}$ to $x_{L,k+1}$.
Of course, in the standard case with $x_c'>0$, the latter corresponds to $S(I_k\cap I_{k+1})$.
However, when $x_c'<0$, the term which we subtract off is actually  $S(I_k\cup I_{k+1})$!
If we extend these observations to a general family of intervals $\lbrace I_k\rbrace$, the bound
on the BH entropy of the outer envelope becomes
 \be
\hS(\lbrace I_k\rbrace)\leq \sum S(I_k)-\sum_{x_{c,k}<x_{c,k+1}}
S(I_k\cap I_{k+1})-\sum_{x_{c,k}>x_{c,k+1}}
S(I_k\cup I_{k+1})\,,
\labell{post5}
 \ee
where $x_{c,k}$ denotes the center of the interval $I_k$. The right-hand
side of this expression replaces eq.~\reef{residue} as our definition of the
\name entropy
 \be
E\equiv\sum_{k=1}^n S(I_k)-\sum_{x_{c,k}<x_{c,k+1}}
S(I_k\cap I_{k+1})-\sum_{x_{c,k}>x_{c,k+1}}
S(I_k\cup I_{k+1})\,.
 \labell{residue2}
 \ee

\begin{figure}[h!]
\begin{center}
\includegraphics[width=0.7\textwidth]{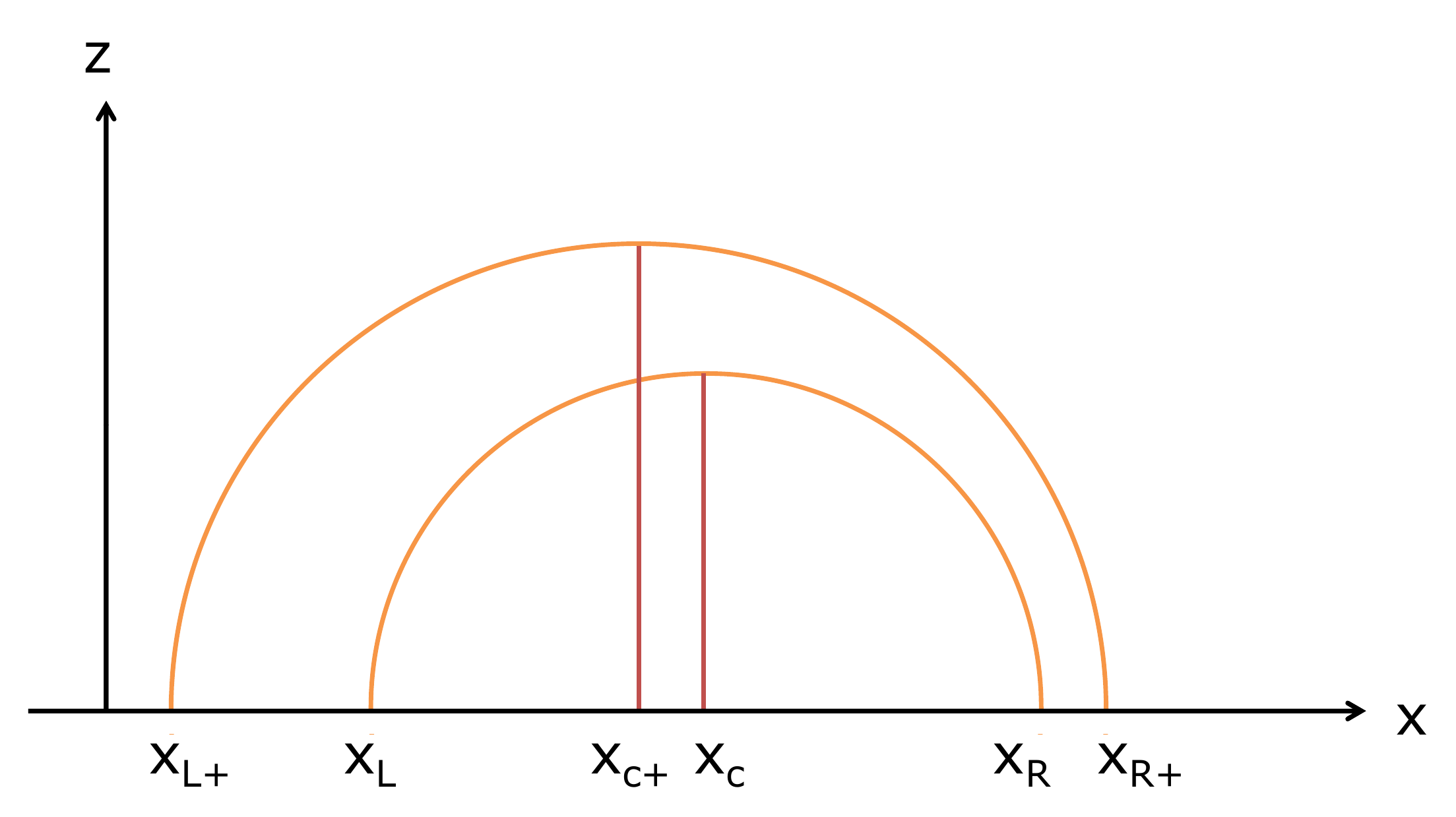}
\caption{(Color online) No intersection between two neighboring tangent semicircles.}
\label{fig8}
\end{center}
\end{figure}

As a final geometric subtlety, let us consider whether or not we will ever encounter
with our construction, the situation illustrated in figure \ref{fig8} where one interval
is completely enclosed by the next
interval in the sequence, \ie either $I_k\subset I_{k+1}$ or $I_k\supset I_{k+1}$.
If such a situation occurred, it would of course produce a problem in defining the outer envelope, as
the dual semi-circles in the bulk would not intersect anywhere. However,
we can show that, in fact, this situation never arises in the continuum limit by showing that
$x_L'\,x_R'\ge0$ everywhere. Note that $x_L(x)=x_c(x)- r(x)$ and $x_R(x)=x_c(x)+ r(x)$ and hence
the desired inequality is
 \be
x_c'^2-r'^2\ge 0\,.
 \labell{requ0}
 \ee
Recall that the derivative of the center point is given in eq.~\reef{movec}. Further with $r=\sqrt{z^2+a^2}$,
one can easily show that
 \be
r'= \frac{z'\,x_c'}{\sqrt{1+z'^2}}
 \labell{inter}
 \ee
and hence, as desired,
 \be
x_c'^2-r'^2=\frac{x_c'^2}{1+z'^2}\ge 0\,.
 \labell{requ0}
 \ee
%

\section{Planar holes in higher dimensions}
\labell{higher}

\begin{figure}[h!]
\begin{center}
\includegraphics[width=0.8\textwidth]{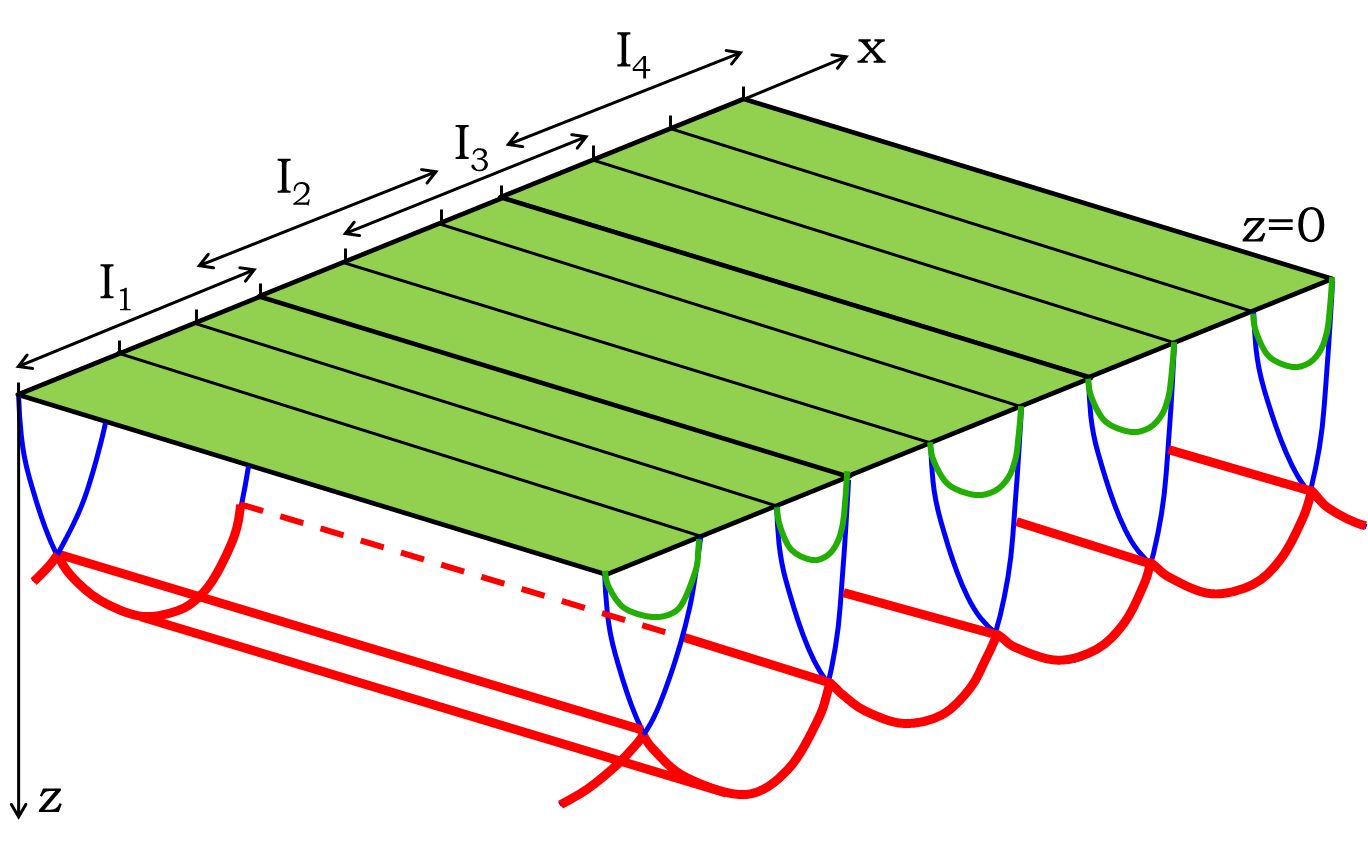}
\caption{(Color online) Covering the boundary of AdS$_{d+1}$ with
a family of overlapping strips.}
\label{slabs}
\end{center}
\end{figure}

We would like to explore whether the construction discussed in the
previous section for a three-dimensional AdS bulk extends to
the case of higher dimensions. As before, we will consider the case
where the $d$-dimensional boundary geometry is simply
flat space. Hence, we are working
in Poincar\'e coordinates with
\beq
ds^2 = \frac{L^2}{z^2}\(dz^2-dt^2+d\vec{x}^2\)\,,
\labell{point}
\eeq
where as usual, the AdS boundary is at $z=0$.  As a simple first step, we will limit
ourselves to considering the case where a
constant time slice is partitioned by a family of overlapping strips or slabs, $\lbrace I_k
\rbrace$, as shown in Figure \ref{slabs}. In general, we will allow the width of the strips
to vary as we move along the orthogonal $x$-axis. Hence our intuition at this stage is
that our construction will allow us to evaluate the Bekenstein-Hawking entropy for bulk
surfaces with a planar symmetry. That is, we can accommodate bulk surfaces with a profile of the
form $z=z(x)$. In order to regulate the area of these surfaces, we will impose that the $x$
direction is periodic with period $\ell_1$, as in section \ref{hole3}. Further, for simplicity,
we also assume that the remaining spatial directions $x^i$ are periodic on a scale $\ell_i$
(for $i=2,\cdots,d-1$, while $i=1$ denotes $x$), in order to regulate the distances along the strips.

As the geometry of the boundary regions and their intersections are both strips, let us
begin by recalling the result for the holographic entanglement entropy of a strip of
width $\Delta x$ on the boundary of AdS$_{d+1}$ \cite{rt1} --- see also \cite{gb}
 \be
S(\Delta x)=\frac{L^{d-1}}{4\Gn}\,\frac{\ell_2\cdots \ell_{d-1}}{d-2}\(\frac{2}{\delta^{d-2}}
-\frac{c_d^{d-1}}{\Delta x^{d-2}}\)\,. \labell{14}
 \ee
Here,
$\delta$ is the usual short-distance cut-off in the boundary CFT and $c_d$ is a numerical constant
given by
 \be
c_d=2\sqrt{\pi}\,\frac{\Gamma\(\frac{d}{2d-2}\)}{\Gamma\(\frac{1}{2d-2}\)}=\frac{\Delta x}{z_*}\,.
 \labell{6}
 \ee
As noted above $c_d$ is also the ratio between the width $\Delta x$ of the boundary interval and
the corresponding maximal height $z_*$ of the extremal surface in the bulk, which is used to
evaluate the holographic entanglement entropy. Recall the logarithmic result \reef{cc3} for the entanglement
entropy in $d=2$ and so implicitly, we are assuming that $d\ge 3$ above. In this case, the
profile of the extremal surface along $x$ direction is no longer
a semicircle but rather a curve in ($z$,$x$)-plane governed by the differential equation \cite{rt1}
 \be
\frac{dz}{dx}=\pm\left[\left(\frac{z_*}{z}\right)^{2d-2}-1\,\right]^{1/2}\,.
 \labell{2}
 \ee

To begin, we consider a bulk surface with a constant profile, $z=z_*$ (and $t=0$).
With the relevant geometry described above, the Bekenstein-Hawking entropy for this surface is given by
 \be
 \frac{\A(z=z_*)}{4\,\Gn}=\frac{L^{d-1}\, \ell_1\ell_2\cdots\ell_{d-1}}{4\Gn\,z_*^{d-1}}\,.
\ee
Now as in the previous section, we expect that this result will be reproduced by evaluating the
\name entropy for a series of $n$ equally spaced strips $I_k$ with a fixed width $\Delta x$
and taking the continuum limit $n\to\infty$. In order that the extremal surfaces determining
the holographic entanglement entropy of the individual strips are tangent
to the desired bulk surface, we must choose $\Delta x=c_d\,z_*$ according to eq.~\reef{6}.
With $n$ strips equally spaced along the $x$ direction, which has a length $\ell_1$, the width
of the intersections $I_k\cap I_{k+1}$ is simply
 \be
w_\textrm{in}=\Delta x-\frac{\ell_1}{n}\,.
 \labell{win2}
 \ee
Then the desired \name entropy \reef{residue2} becomes
 \beqa
E&=&\lim_{n\to\infty} n(S(\Delta x)-S(w_\textrm{in}))\nonumber\\
&=& \frac{L^{d-1}}{4\Gn}\,\frac{\ell_2\cdots\ell_{d-1}}{d-2}\,c^{d-1}_d\,
\lim_{n\to\infty}n\(-\frac{1}{\Delta x^{d-2}}+\frac{1}{(\Delta x-\ell_1/n)^{d-2}}\)
\nonumber\\
&=& \frac{L^{d-1}}{4\Gn}\,\frac{\ell_2\cdots\ell_{d-1}}{d-2}\,c^{d-1}_d\,
\lim_{n\to\infty}\(\frac{(d-2)\ell_1}{\Delta x^{d-1}}+O(1/n)\)
\nonumber\\
&=&\frac{L^{d-1}\,\ell_1\ell_2\cdots\ell_n}{4\Gn\ z_*^{d-1}}\,. \labell{8}
 \eeqa
Hence we see that that the construction in section \ref{hole3} naturally extends to higher
dimensions, at least for the case of a bulk surface with a constant profile.

\begin{figure}[h!]
\begin{center}
\includegraphics[width=0.4\textwidth]{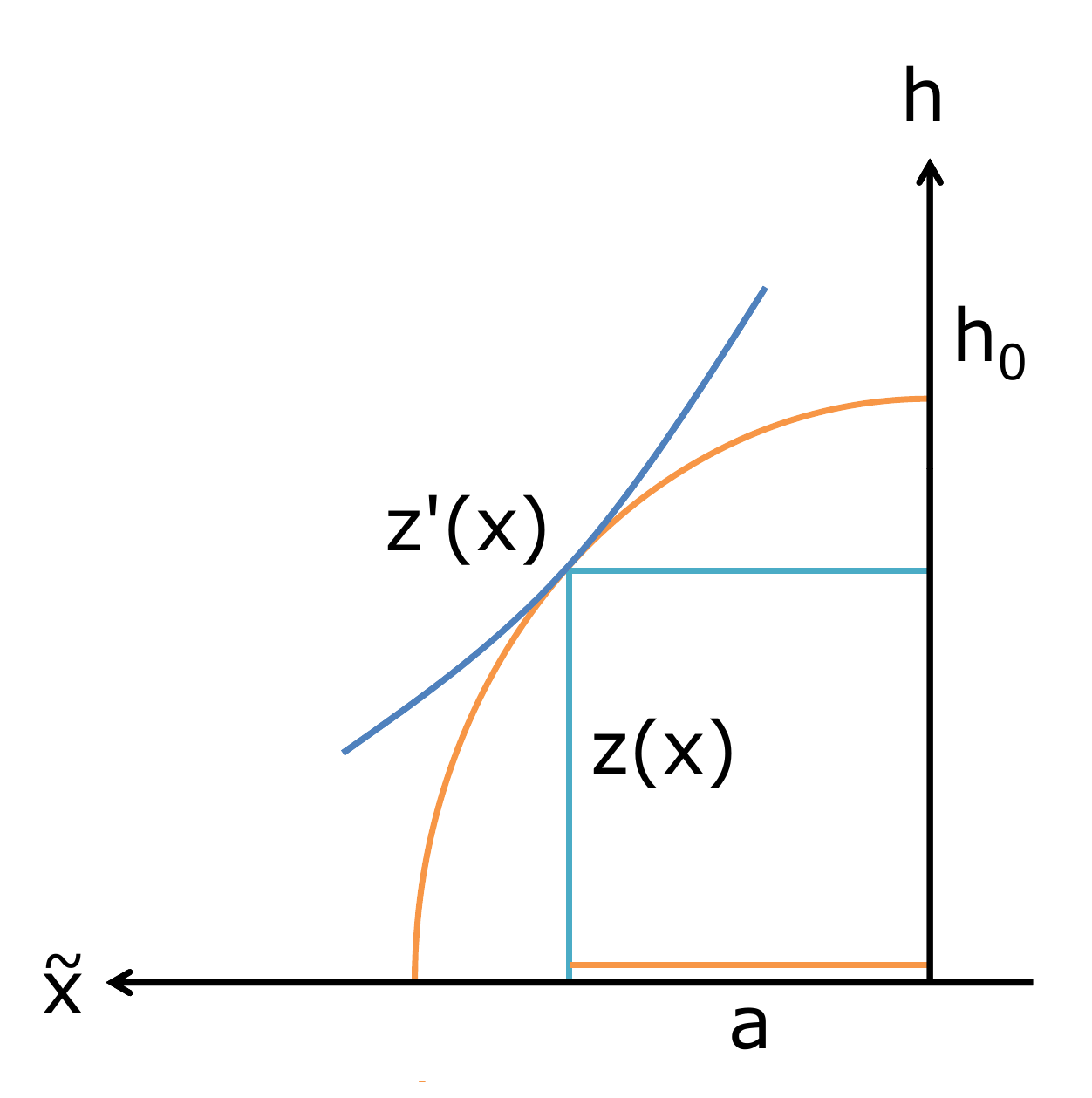}
\caption{(Color online) A varying $z$ profile and its local tangent surface in AdS$_{d+1}$.}
\label{fig5}
\end{center}
\end{figure}
Given this success, we move to considering a bulk surface of nontrivial profile $z(x)$, \ie still respecting
the planar symmetry. Evaluating the BH formula \reef{prop0} on such a surface
in AdS$_{d+1}$ yields
 \be
 \frac{\A(z=z(x))}{4\,\Gn}=\frac{L^{d-1}}{4\Gn}\,\ell_2\cdots\ell_{d-1}\,
\int_0^{\ell_1}\!dx\,\frac{\sqrt{1+z'^2}}{z^{d-1}}\,.
 \labell{goal2}
\ee
Now the first step towards evaluating the \name entropy \reef{residue2} will
be identifying the extremal surface which is tangent to the bulk surface at a given $x$,
as shown in figure \ref{fig5}. We denote the profile of these extremal surfaces as
$h(\tilde x; x)$, where the second argument indicates that this extremal profile
is tangent to the bulk surface at $\tilde x=x$, \ie
\be
h(\tilde x=x;x)=z(x)\,, \qquad \frac{dh(\tilde x; x)}{d\tilde x}\bigg|_{\tilde x=x}=z'(x)
\,. \labell{4}
 \ee
Further, with eq.~\eqref{2}, we can write
 \be
\frac{dh}{d\tilde x}=\left[\left(\frac{h_0}{h}\right)^{2d-2}-1\,\right]^{1/2}\,,
 \labell{5}
 \ee
where $h_0(x)$ is the maximal height to which the extremal profile $h(\tilde x; x)$ rises in the bulk.
Note that we have implicitly assumed that  $dh/d\tilde x\ge0$ above. In this case,
as shown in figure \ref{fig5}, we consider the shift $a(x)$
along the $x$-axis between the tangent point $x$ and the midpoint of the boundary interval
$x_c(x)$, where the extremal surface reaches $h_0$. This quantity is determined by
 \be
a(x)=\int_x^{x_c}d\tilde x=\int_{z(x)}^{h_0}\frac{h^{d-1}dh}{\sqrt{h_0^{2d-2}-h^{2d-2}}}
=\frac{h_0}{2d-2}\,B\!\left[\(\frac{z}{h_0}\)^{2d-2}\right]\,,
 \labell{shift88}
 \ee
where
 \be
B[x]=\int_x^1\frac{ds}{s^{\frac{d-2}{2d-2}}\sqrt{1-s}}\,.
 \labell{bee}
 \ee
On the other hand, combining eqs.~\eqref{4} and \eqref{5} yields
 \be
z'^2=\frac{h_0^{2d-2}}{z^{2d-2}}-1\,,
 \ee
and therefore we may write
 \be
h_0=z(1+z'^2)^{1/(2d-2)}\,,\qquad
a=\frac{h_0}{2d-2}\,B\!\left[\frac{1}{1+z'^2}\right]\,.
 \labell{bumpy}
 \ee
Further, according to eq.~\eqref{6}, the width of the interval is
 \be
\Delta x=c_d\,h_0=c_d\,z\,(1+z'^2)^{1/(2d-2)}\,.
 \ee

Having established these preparatory results, we now consider the intervals,
$I_{k-1}$, $I_k$, $I_{k+1}$, for which the extremal bulk surfaces are tangent
to the profile $z(x)$ at $x-dx$, $x$ and $x+dx$, respectively. Now as in eq.~\reef{over2},
we denote the
width of the intersections $I_{k-1}\cap I_k$ and $I_{k}\cap I_{k+1}$, respectively,
as
 \beqa
o_\pm&=&\frac{1}{2}(\Delta x(x)+\Delta x(x\pm dx))\pm(a(x)-a(x\pm dx))-dx
\nonumber\\
&=&\Delta x -\(1+a' \mp\Delta x'\)\,dx\,,
 \labell{file9}
 \eeqa
where, to leading order in $dx$, we have used
 \be
\Delta x(x\pm dx)=\Delta x(x)\pm\Delta x'\,dx\,,\qquad
a(x\pm dx)=a(x)\pm a'\,dx\,.
 \ee
From eq.~\reef{bee}, we have
 \be
\partial_xB\!\[(z/h_0)^{2d-2}\]=-\frac{1}{s^{\frac{d-2}{2d-2}}\sqrt{1-s}}\frac{ds}{dx}\bigg|_{s=1/(1+z'^2)}
=\frac{2z''}{(1+z'^2)^{1+1/(2d-2)}}\,, \labell{12}
 \ee
and hence we may write
 \be
a'=\frac{1}{2d-2}\(B\,h_0'+\frac{2zz''}{1+z'^2}\)\,.
 \ee
Hence we can re-express the overlaps in eq.~\reef{file9} as
 \be
o_\pm=\Delta x-\[1+\frac{1}{2d-2}\(B\,h_0'+\frac{2zz''}{1+z'^2}\)
\mp\Delta x'\]dx\,.
 \labell{over99}
 \ee

Now if we use an `averaged' expression for the \name entropy, as in eq.~\reef{residue2}, we find
 \beqa
E&=&\frac{L^{d-1}}{4\Gn}\,\frac{\ell_2\cdots\ell_{d-1}}{d-2}\,c_d^{d-1}
\int_0^{\ell_1}dx\,\(-\frac{1}{\Delta x^{d-2}}+\frac{1}{2\,o_+^{d-2}}
+\frac{1}{2\,o_-^{d-2}}\)
\nonumber\\
&=&\frac{L^{d-1}}{4\Gn}\,\frac{\ell_2\cdots\ell_{d-1}}{d-2}\,c_d^{d-1}
\int_0^{\ell_1}dx\,\frac{1}{\Delta x^{d-1}}\(\Delta x-\frac{o_++o_-}{2}\)\,,
\nonumber\\
&=&\frac{L^{d-1}}{4\Gn}\,\frac{\ell_2\cdots\ell_{d-1}}{d-2}\,
\int_0^{\ell_1}dx\,\frac{1}{h_0^{d-1}}
\[1+\frac{1}{2d-2}\(B\,h_0'+\frac{2zz''}{1+z'^2}\)\]\,. \labell{10}
 \eeqa
where we used eq.~\reef{bumpy} to replace $h_0$ in the final line.
Unfortunately, at this stage, the above expression looks quite different from the desired result \reef{goal2}.
However, given our discussion in the previous section, we should expect that the integrands in
these two expressions will differ by a total derivative.
Hence we examine the difference between the two
integrands and applying eqs.~\reef{bumpy} and \reef{12}, one finds
 \beqa
&&\frac{1}{h_0^{d-1}}
\[1+\frac{1}{2d-2}\(B\,h_0'+\frac{2zz''}{1+z'^2}\)\]-\frac{\sqrt{1+z'^2}}{z^{d-1}}
\nonumber\\
&&\quad
=\frac{d}{dx}\(-\frac{1}{2(d-1)(d-2)}\frac{B}{h_0^{d-2}}+\frac{1}{d-2}\frac{z'}{z^{d-2}\sqrt{1+z'^2}}\)\,.
 \labell{totalD}
 \eeqa
The contribution of this total derivative vanishes as a boundary term and hence
we have confirmed that the \name entropy again yields the BH entropy \reef{prop0}
for these bulk surfaces with a nontrivial profile $z(x)$.

\subsection{Higher dimensions and higher curvatures}
\labell{gbgra}

At this point, we would like to comment on extending these calculations to theories
of higher curvature gravity in the bulk. In particular, we have shown that this discussion
can accommodate Gauss-Bonnet gravity, in which a curvature-squared interaction proportional
to the four-dimensional Euler density is included in the action. In a holographic context, this
theory is often considered as a toy model to describe boundary CFT's where the central charges
are not all equal \cite{gb,gb2}.
As was first discussed in \cite{highc,highc2}, holographic entanglement entropy can be
calculated with a simple extension of the Ryu-Takayanagi prescription.
In particular, one replaces the BH formula in eq.~\reef{define} with the
following entropy functional
  \be
S_\mt{JM}=\frac{1}{4\Gn}\int_\sigma d^{d-1}x\sqrt{h}\(1+\frac{2\la L^2}{(d-2)(d-3)}{\cal R}\)
 \labell{21}
 \ee
where $\cal R$ is the intrinsic curvature scalar for the bulk surface $\sigma$ and
$\lambda$ is the (dimensionless) coupling for the curvature-squared terms in the action
--- see appendix \ref{loveA}.\footnote{Note that we have dropped a surface term
that should naturally be included here \cite{highc} as it will be irrelevant for our discussion.}
This entropy functional was originally derived in studying black hole
entropy for these theories \cite{tedJ}. Nontrivial tests of holographic entanglement entropy were
made with this prescription in \cite{highc,highc2}, however, following \cite{aitor}, this can now
be derived \cite{xifriend}.

At this point, we simply re-iterate that one is able to extend the previous discussion to
incorporate these theories. In particular, the final result is that in the continuum limit,
the \name entropy in the boundary theory, which is evaluated holographically using $S_\mt{JM}$, matches the
gravitational entropy in the bulk, which in this case is given by evaluating $S_\mt{JM}$ on the bulk surface.
The proof of this statement using the approach of the present section is rather lengthy and tedious. Hence
we do not provide the details here and rather we note that this result is a corollary of the general proof
appearing in section \ref{general}.

\begin{figure}[h!]
\begin{center}
\includegraphics[width=0.4\textwidth]{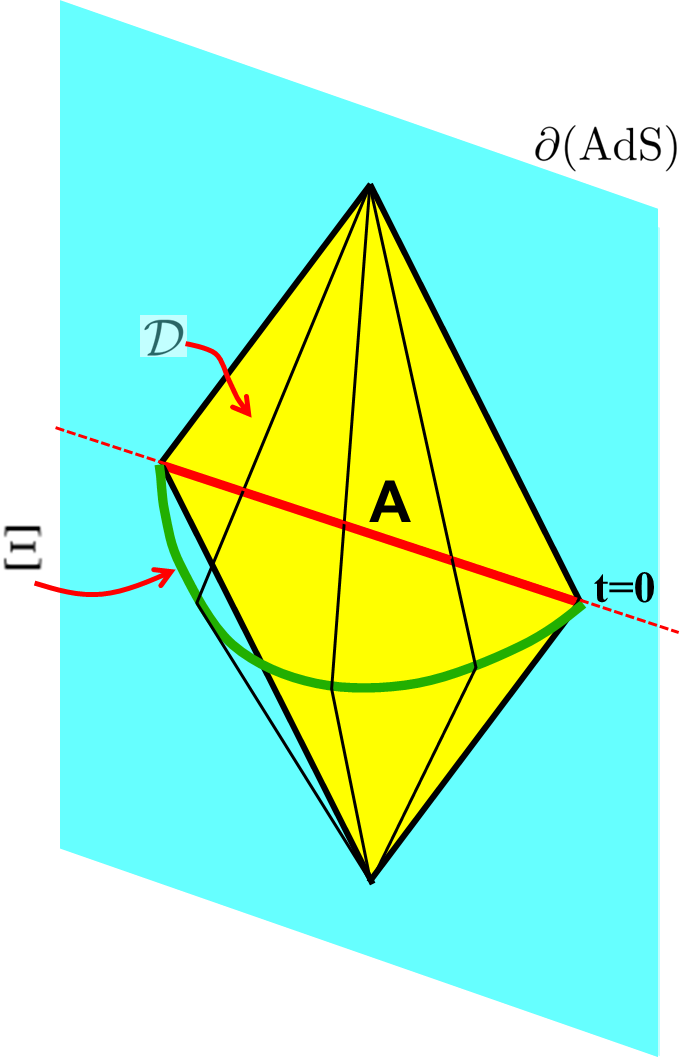}
\caption{(Color online) The geometry relevant for the construction of causal holographic information
--- see the discussion in the main text.}
\label{wedge}
\end{center}
\end{figure}
%

\section{Causal holographic information}
\labell{cause}

Causal holographic information has been conjectured to be another interesting
measure of entanglement in the boundary theory in a holographic framework
\cite{causal0,aron}. As we will comment below, it also has a natural connection
to the discussion of holographic holes in \cite{hole}. Hence let us review
the definition of causal holographic information --- see figure \ref{wedge}:
 One begins by specifying a region $A$
on a Cauchy surface in the boundary theory. One constructs the causal development
$\cal D$ of this region, again in the boundary, and then extends null rays into the
bulk from the boundary of $\cal D$ --- past-directed light rays from the future
boundary $\partial{\cal D}^+$ and future-directed light rays from the past
boundary $\partial{\cal D}^-$. The envelope of these null rays enclose a bulk
region, known as the causal wedge of $A$. The causal holographic information
is then defined by evaluating the Bekenstein-Hawking entropy on the extremal
surface on the boundary of the causal wedge, \ie
\beq
\chi(A)= {\rm ext}\,\frac{{\cal A}(\Xi)}{4\Gn}
\labell{cause0}
\eeq
where as in the holographic entanglement entropy, one extremizes over surfaces
$\Xi$ which are homologous to $A$, but now confined to the boundary of the
causal wedge.

Generally, the holographic entanglement entropy and the causal holographic
information are distinct quantities. In particular, the extremal surface used to
evaluate the holographic entanglement entropy of a given region
typically probes deeper into the bulk than that appearing
in the causal holographic information. An exception to this generic behavior arises with a spherical entangling
surface and the boundary CFT in its vacuum state, \ie the bulk is described by the
pure AdS$_{d+1}$ vacuum. In this case, the corresponding causal wedge corresponds
to an AdS-Rindler patch in the bulk and the extremal surface selected with the
RT prescription \reef{define} is precisely the bifurcation surface of the corresponding
AdS-Rindler horizon \cite{chm,eom1}. Hence the two extremal surfaces precisely
match.\footnote{We note that this match extends to the case where the bulk
theory is described by a classical gravity theory with any arbitrary higher curvature
action \cite{chm,eom1}.} Hence for a single interval in a two-dimensional boundary
CFT, \ie an AdS$_3$ bulk, the causal holographic information generally matches the holographic entanglement entropy.\footnote{We
thank Veronika Hubeny for emphasizing that this matching will be violated in
certain special cases even with $d=2$, \eg for large intervals in a thermal state \cite{plateau}.}
Hence the analysis of \cite{hole} does not distinguish between these two quantities. In
fact, this special feature is an essential part of the discussion in \cite{hole}, since the
contribution of each interval in eq.~\reef{residue} is associated with the information
which an accelerated bulk observer in the associated causal wedge can collect. Hence
natural extension of the discussion in \cite{hole} to higher dimensions might seem
to involve constructing the surfaces in the bulk using the extremal surfaces
used to evaluate the causal holographic information.

Hence we examine a version of our construction in the previous section using the
causal holographic information. That is,  we replace
the entanglement entropies in the \name entropy \reef{residue} with the corresponding
causal holographic information for the same intervals to define the
`\name causal holographic information,'
 \be
E_\chi\equiv\sum_{k=1}^n\chi(I_k)-\sum_{k=1}^n\chi(I_k\cap I_{k+1})
\,. \labell{reschi}
 \ee
Note that the causal holographic information does not satisfy the equivalent
of strong subadditivity \reef{ssa1} and so this motivation is lacking when
we apply eq.~\reef{reschi}.

If we are considering a strip of width $\Delta x$ on the boundary of
AdS$_{d+1}$, as in section \ref{higher}, the corresponding extremal
surface which defines the causal holographic information is a half cylinder defined by
 \be
z^2+(x-x_c)^2=(\Delta x/2)^2\,.
\label{chi9}
 \ee
Given the causal holographic information is then given by evaluating the area of this surface, namely
 \beqa
\chi_d&=&\frac{L^{d-1}}{4\Gn}\,\ell_2\cdots\ell_{d-1}\int\frac{dx}{z^{d-1}}
\sqrt{1+z'^2}\nonumber\\
&=& \frac{L^{d-1}}{2\Gn}\,\ell_2\cdots\ell_{d-1}\int_\delta^{\Delta x/2}\frac{dz}{z^{d-1}}
\(1-\frac{4\,z^2}{\Delta x^2}\)^{-1/2}\,.
 \labell{chi8}
 \eeqa
Now it is straightforward to evaluate the above integral of a given value of $d$. However, we will be
primarily interested in the leading singularities as $\delta\to0$ and so we approximate the integral
as
 \beqa
 f_d(\Delta x)&\equiv&\int_\delta^{\Delta x/2}\frac{dz}{z^{d-1}}
\(1-\frac{4\,z^2}{\Delta x^2}\)^{-1/2}
\nonumber\\
&\simeq&\int_\delta^{\Delta x/2}\frac{dz}{z^{d-1}}
\(1+\frac{2\,z^2}{\Delta x^2}+\cdots\)
\nonumber\\
&\simeq&\frac{1}{(d-2)\,\delta^{d-2}}+\frac{2}{(d-4)\,\Delta x^2\,\delta^{d-4}}+\cdots\,.
\labell{fff9}
 \eeqa

Now we wish to see if eq.~\reef{reschi} can be used to reproduce the BH
entropy \reef{prop0} evaluated for closed surfaces in the bulk. For simplicity, we restrict our attention
to bulk surfaces with a constant profile, \ie $z=z_*$. Then
following the approach in the previous section, we wish to evaluate eq.~\reef{reschi}
for a series of $n$ equally spaced strips with a fixed width $\Delta x=2z_*$
and then to take the continuum limit $n\to\infty$. This yields
 \beqa
E_\chi&=&\lim_{n\to\infty} n\,(\chi(\Delta x)-\chi(\Delta x-\ell_1/n))
\nonumber\\
&=&\frac{L^{d-1}}{2\Gn}\,\ell_2\cdots\ell_{d-1}\,
 \lim_{n\to\infty} n(f_d(\Delta x)-f_d(\Delta x-\ell_1/n))
\nonumber\\
&=&-\frac{2\,L^{d-1}}{(d-4)\Gn}\,\frac{\ell_1\ell_2\cdots\ell_{d-1}}{\Delta x^3\,\delta^{d-4}}
=-\frac{1}{d-4}\(\frac{z_*}{\delta}\)^{d-4}\,\frac{{\cal A}(z=z_*)}{4\,\Gn}\,.
 \labell{chi789}
 \eeqa
Hence we see that the \name causal holographic information does not match the BH entropy of
the bulk surface.

In fact, the result differs from the BH entropy by a factor
which diverges in the limit $\delta\to0$. It is not hard to understand the origin of this divergence.
Just as with the entanglement entropy, the causal holographic information contains a number of power law divergences,
as illustrated in eq.~\reef{fff9}. The leading singularity yields the usual area law term, however, the
coefficients of subleading divergences are nonlocal and in general depend on the entire geometry
of the entangling surface \cite{ben}. Hence in the differences appearing in the \name causal holographic information \reef{reschi}, the
area law divergences cancel but the subleading divergences to not because of their nonlocal character.
We stress that the coefficients of all of the power law divergences appearing in the entanglement entropy
can be expressed as local integrals of various geometric factors over the entangling surface \cite{calc}.
Hence, we can generally expect that these divergences will cancel in differences of entanglement entropies,
as long as the same boundaries appear in the positive and negative contributions.

To close, we note that there are two special cases (with $d>2$) where the result in eq.~\reef{chi789}
does not apply, \ie $d=3$ and 4. In those cases, one finds
 \beqa
d=4:&&E_\chi=-\log\(\frac{z_*}{\delta}\)\ \frac{{\cal A}(z=z_*)}{4\,\Gn}\,.
\nonumber\\
d=3:&&E_\chi=\frac{\delta}{z_*}\ \frac{{\cal A}(z=z_*)}{4\,\Gn}\,.
 \labell{special}
 \eeqa
Hence rather than a power law divergence, the calculation in $d=4$ yields a logarithmic divergence,
as should have been expected. In contrast, for $d=3$, the result will vanish in the limit $\delta\to0$
rather than diverging.

\section{General holographic backgrounds}
\labell{general}

In this section, we will show that the anti-de Sitter background was not an essential
ingredient for the agreement in section \ref{higher}. Rather the matching between
the \name entropy in the boundary theory and the gravitational entropy of surfaces
in the bulk (in the continuum limit) is a result that extends to a general holographic
framework. The only essential assumption will be that the entanglement entropy in the boundary theory
is still calculated holographically by the Ryu-Takayanagi prescription \reef{define}. In fact, in
the last part of this section, we will extend to the discussion to more general entropy functionals.
In particular, the general form considered there will accommodate the holographic prescription for
calculating entanglement entropy where the bulk is described by Lovelock gravity \cite{highc,highc2}.

To begin, we consider the following general metric to describe our
($d$+1)-dimensional holographic background:
 \be
ds^2=-g_0(z)dt^2+\sum_{i=1}^{d-1}g_i(z)(dx^i)^2+g_1(z)f(z)dz^2\,.
\labell{back}
 \ee
This background geometry should arise as the solution of some classical
gravity equations, perhaps with some background fields, but the details
of these equations will be unimportant for our considerations.
As usual, we will assume that the asymptotic boundary is reached with the limit
$z\to0$. As usual to regulate the area of the surfaces considered below, we will
assume that the spatial coordinates $x^i$ are periodic with some large period
$\ell_i$. In particular, we choose a surface in the bulk with a profile $z=z(x)$
(where $x=x^1$, as before)
and so which respects the planar symmetry introduced in section \ref{higher}.
The Bekenstein-Hawking entropy of this surface is then given by
 \be
 \frac{\A(z=z(x))}{4\,\Gn}=\frac{\ell_2\cdots\ell_{d-1}}{4\Gn}\,
\int_0^{\ell_1}\!dx\,\sqrt{G(z)}\sqrt{1+f(z)\,z'^2}\quad{\rm where}\ \
G(z)=g_1\cdots g_{d-1}\,.
 \labell{goal77}
 \ee
Now our goal is to show that we can reproduce this expression using the \name
entropy \reef{residue2}.

For simplicity, we begin by considering a bulk surface with the constant profile $z=z_*$.
In this case, eq.~\reef{goal77} reduces to
 \be
 \frac{\A(z=z_*)}{4\,\Gn}=\frac{\ell_1\ell_2\cdots\ell_{d-1}}{4\Gn}\,
\sqrt{G_*}\qquad{\rm where}\ \
G_*=G(z_*)\,.
 \labell{goal88}
 \ee
Following the discussion in section \ref{higher}, we would like to reproduce this
result using the \name entropy applied to a family of strips in the boundary
equally spaced along the $x$ direction and each with the same width
$\Delta x$.

As usual, the holographic entanglement entropy of a strip
will be determined by an extremal surface with a profile respecting the planar
symmetry of the geometry, \ie $z=h(x)$. Evaluating eq.~\reef{define} in the present framework
then yields
 \be
S(\Delta x)=\frac{\ell_2\ldots\ell_{d-1}}{4\Gn}\,\sigma(\Delta x)\,,
 \ee
where
 \be
\sigma(\Delta x)=\int_0^{\Delta x} dx\sqrt{G(h)}\sqrt{1+f(h)\,h'^2}\,. \labell{38}
 \ee
We will now go through a series of steps to show that $d\sigma/d\Delta x$ has a particularly
simple form. The latter will then be useful in showing that the bulk gravitational entropy
matches the \name entropy in the boundary theory.

Treating eq.~\reef{38} as an effective action, there is a conserved `energy' because the integrand
has no explicit $x$ dependence. The conserved quantity can be written as
 \be
\frac{\sqrt{G(h)}}{\sqrt{1+f(h)\,h'^2}}=\sqrt{G_0}\qquad
{\rm where}\ \ G_0=G(h_0)\,,
\labell{energy}
 \ee
and where $h_0$ is the maximal value of the profile, where $h'=0$.
Eq.~\reef{energy} can be re-expressed as a first-order equation of motion for the
extremal profile,
 \be
h'=\pm\[\frac{G(h)-G_0}{G_0\,f(h)}\]^{1/2}\,. \labell{39}
 \ee
Now we change the integration variable in eq.~\reef{38} from $x$ to $h$,
 \be
\sigma(\Delta x)=2\int_\delta^{h_0}\frac{dh}{h'}\sqrt{G(h)}\sqrt{1+f(h)\,h'^2}\,,
 \ee
where implicitly we are only integrating over the half of the extremal surface on which $h'\ge0$.
We have also introduced a short-distance cut-off $\delta$ to regulate any UV divergences in
the entanglement entropy arising from $h\to0$. Next we can
eliminate $h'$ using eq.~\eqref{39}, which yields
 \be
\sigma(\Delta x)=2\int_\delta^{h_0}dh\frac{\sqrt{f(h)}\,G(h)}{\sqrt{G(h)-G_0}}\,.
 \ee
We can also produce a similar expression for the width of the strip,
 \be
\Delta x=2\int_\delta^{h_0}\frac{dh}{h'}=2\sqrt{G_0}\int_\delta^{h_0}dh\,
\frac{\sqrt{f(h)}}{\sqrt{G(h)-G_0}}\,.
 \labell{spread}
 \ee
Combining these two equations above, one can show that
 \be
\sigma(\Delta x)=\sqrt{G_0}\,\Delta x+2\int_\delta^{h_0}dh\,\sqrt{f(h)}\,\sqrt{G(h)-G_0}\,.
 \ee
Now we differentiate this last expression with respect to $h_0$ to find
 \be
\frac{d\sigma}{dh_0}=\sqrt{G_0}\,\frac{d\Delta x}{dh_0}+
\frac{1}{2\sqrt{G_0}}\frac{dG_0}{dh_0}\Delta x+
2\sqrt{f\,(G-G_0)}\big|_{h=h_0}
-\frac{dG_0}{dh_0}
\int_\delta^{h_0}dh\frac{\sqrt{f}}{\sqrt{G-G_0}}\,.
 \ee
However, the right-hand side above can be greatly simplified. First, the third term vanishes
because $G(h=h_0)=G_0$. Second, from eq.~\reef{spread}, we can recognize the integral
in the fourth term yields $\Delta x/(2\sqrt{G_0})$. With this substitution, the second
and fourth terms cancel and we are left with
 \be
\frac{d\sigma}{dh_0}=\sqrt{G_0}\,\frac{d\Delta x}{dh_0}\,.
 \labell{square}
 \ee
Alternatively, we can write
 \be
\frac{d\sigma}{d\Delta x}=\frac{d\sigma}{dh_0}\bigg/\frac{d\Delta x}{dh_0}=\sqrt{G_0}\,.
 \labell{usefulx}
 \ee
Note that this is a general result for the strip entropy, that is independent of our choice of a bulk surface.

Now following the discussion of section \ref{higher}, the proof that the \name entropy
matches eq.~\reef{goal88} is straightforward. In particular, we have $n$ intervals of a fixed
width $\Delta x$ equally spaced along the $x$ direction. The width $\Delta x$ will be chosen
so that the extremal surfaces touch the bulk surface at their maxima, \ie $h_0=z_*$. Then
in parallel with eq.~\reef{8}, the desired \name entropy becomes
 \beqa
E&=&\lim_{n\to\infty} n\(S(\Delta x)-S\(\Delta x-\frac{\ell_1}n\)\,\)\nonumber\\
&=& \frac{\ell_2\cdots\ell_{d-1}}{4\Gn}\,
\lim_{n\to\infty} n\(\sigma(\Delta x)-\sigma\(\Delta x-\frac{\ell_1}n\)\,\)\nonumber\\
\nonumber\\
&=& \frac{\ell_2\cdots\ell_{d-1}}{4\Gn}\,
\lim_{n\to\infty}\(\ell_1\,\frac{d\sigma}{d\Delta x}+O(1/n)\)
\nonumber\\
&=&\frac{\ell_1\ell_2\cdots\ell_{d-1}}{4\Gn}\,\sqrt{G_*}\,, \labell{8g}
 \eeqa
where in the last line, we have used $G_*=G_0$. Hence the \name entropy precisely reproduces
eq.~\reef{goal88} in the continuum limit.

Now we would like to reproduce the general expression \reef{goal77} for a bulk surface
with a nontrivial profile $z=z(x)$. In this case, the family of strips will be chosen on the boundary so that
there is a dual extremal surface tangent to each point on this profile. That is, as in section \ref{higher},
we have a family of extremal surfaces $z=h(\tilde x; x)$, which are chosen to satisfy the two conditions
in eq.~\reef{4}. Hence, the width of the strips becomes a function of the position of the
tangent point along the bulk curve.
The general expression for the
width of the intersection of neighboring strips given in eq.~\reef{file9} will still apply in the
present situation. Hence in the `averaged' expression for the \name entropy \reef{residue2}, we encounter
 \beqa
&&S(\Delta x)-\frac12\(S(o_+)+S(o_-)\)
\nonumber\\
&&\quad=\frac{\ell_2\cdots\ell_{d-1}}{4\Gn}\,
\(\sigma(\Delta x)-\frac12\sigma\(\Delta x-(1+a'-\Delta x')dx\)
-\frac12\sigma\(\Delta x-(1+a'+\Delta x')dx\)\,\)
\nonumber\\
&&\quad=\frac{\ell_2\cdots\ell_{d-1}}{4\Gn}\,\frac{d\sigma}{d\Delta x}\,(1+a')\,dx
=\frac{\ell_2\cdots\ell_{d-1}}{4\Gn}\,\sqrt{G_0}\,(1+a')\,dx\,,
 \labell{differ9}
 \eeqa
where we have used eq.~\reef{usefulx} in the last step. Note that in the present
situation, $h_0$ and hence $G_0$ are both functions of $x$.
Now with the above expression, the desired \name entropy becomes
 \be
E=\frac{\ell_2\cdots\ell_{d-1}}{4\Gn}\,\int_0^{\ell_1}\! dx\,\sqrt{G_0}(1+a')\,.
 \labell{boro}
 \ee
Again, at first sight, this expression is quite dissimilar from eq.~\reef{goal77}. However,
we expect that the integrands will differ by a total derivative.

First we note that using eq.~\reef{4}, we can re-express eq.~\reef{energy} as
 \be
\frac{G(z)}{1+f(z)z'^2}=G(h_0)\,, \labell{40}
 \ee
where the $z(x)$ appearing on the right-hand side corresponds to the profile of the
bulk surface. Further we can express the shift $a$ between the tangent point $x$ and
the midpoint of the corresponding interval $x_c(x)$ as
 \be
a=\int_x^{x_c}d\tilde x=\int_z^{h_0}\frac{dh}{\partial_{\tilde x}h}
=\sqrt{G_0}\int_z^{h_0}dh\frac{\sqrt{f(h)}}{\sqrt{G(h)-G_0}}\,,
 \labell{boro2}
 \ee
where we have used eq.~\reef{39} in the last step.

Now we wish to show that the following corresponds to a total derivative
 \be
\sqrt{G_0}(1+a')-\sqrt{G}\sqrt{1+fz'^2}\,,
 \ee
in order to prove the equivalence of the gravitational entropy \reef{goal77}
in the bulk and the \name entropy \reef{boro} in the boundary.
For this purpose, consider the auxiliary quantity
 \be
A=\int_z^{h_0}dh\sqrt{f(h)}\sqrt{G(h)-G_0}\,,
\labell{aux1}
 \ee
which is readily shown to satisfy
 \be
\frac{dA}{dx}=-fz'^2\sqrt{G_0}-\frac{1}{2}\frac{G_0'}{\sqrt{G_0}}a\,,
 \ee
using eqs.~\eqref{40} and \reef{boro2}. Then with further substitutions
of eq.~\reef{40}, we find
 \beqa
\sqrt{G_0}(1+a')-\sqrt{G}\sqrt{1+fz'^2}&=&\sqrt{G_0}(a'-fz'^2)
\nonumber\\
&=&\sqrt{G_0}\,a'+\frac{1}{2}\frac{G_0'}{\sqrt{G_0}}a+A'
\nonumber\\
&=&\(\sqrt{G_0}\,a+A\)'\,,
 \eeqa
and as expected, this difference is a total derivative. Therefore the desired equivalence
between eqs.~\reef{goal77} and \reef{boro} has been established.


\subsection{Generalized entropy functionals} \labell{gene}

Next we would like to extend the above discussion to consider
slightly more general entropy functionals. We begin, as before, by
focusing our attention on situations with planar symmetry, \ie
we choose a bulk surface with a profile $z=z(x)$ in
a holographic background of the form given in
eq.~\reef{back}. However, after evaluating the entropy functional
on this surface, we will assume that it takes the form
 \be
S_\mt{grav}(z=z(x))= \ell_2\cdots\ell_{d-1}\,
\int_0^{\ell_1}\!dx\ \cL(z,P)\,.
 \labell{goal77x}
 \ee
where $P=z'^2$. That is, the integrand may have a general dependence on $z$ but
only even powers of $z'$ appear (and no higher derivatives appear). This form \reef{goal77x} is sufficiently general to
incorporate the entropy for any of the Lovelock theories --- see appendix \ref{loveA}.
For example, with Gauss-Bonnet gravity, if we evaluate eq.~\reef{21} for a bulk
surface in AdS space, the result takes the form
 \be
S_\mt{JM} =\frac{L^{d-1}\,\ell_2\ldots\ell_{d-1}}{4\Gn\,f_\infty^{(d-1)/2}}\,
\int_0^{\ell_1}\! \frac{dx}{z^{d-1}}\,\(\sqrt{1+z'^2}+2\la f_\infty \frac{z'^2}{\sqrt{1+z'^2}}\) \,,
 \labell{sjm}
 \ee
where $\lambda$ is the dimensionless coupling associated with the curvature-squared interaction
and $f_\infty=(1-\sqrt{1-4\lambda})/(2\lambda)$ --- \eg see \cite{gb2}.

Now our goal is to show that we can reproduce this expression \reef{goal77x} using the \name
entropy \reef{residue2} for a family of strips distributed along the $x$ direction.
We assume that the RT prescription \reef{define} will be generalized
to involve extremizing over a new geometric entropy functional. Then, the holographic
entanglement entropy of a strip will be determined by an extremal surface with a profile
of the form $z=h(x)$ and the final result will take the form
 \be
S(\Delta x)=\ell_2\ldots\ell_{d-1}\,\sigma(\Delta x)\,,
 \ee
where
 \be
\sigma(\Delta x)=\int_0^{\Delta x}\!dx\,\cL(h,P)\,\,. \labell{38x}
 \ee
and $P=h'^2$ here.
Note that the integrand above has precisely the same functional form as in eq.~\reef{goal77x}.
At this stage, we will again show that $d\sigma/d\Delta x$ has a
simple form. The latter will then be applied in establishing the equivalence of the bulk gravitational entropy
\reef{goal77x} and the \name entropy in the boundary theory.

First the conserved quantity associated with the absence of an explicit $x$ dependence in
eq.~\reef{38x} is
 \be
2\frac{\partial \cL}{\partial P}\,P-\cL=-\cL_0\,,
\labell{bana}
 \ee
where $\cL_0=\cL(h_0,0)$ is the integrand evaluated at the maximal height of the extremal profile, which
we denote as $h_0$. Using this expression, we can write for the extremal action
 \be
\sigma=2\int_\delta^{h_0}\frac{dh}{h'}\cL(h,P)=
2\int_\delta^{h_0}\! dh\,\(\frac{\cL_0}{h'}+2\frac{\partial \cL}{\partial P} h'\)\,.
 \ee
Similarly, the width of the strip can be expressed as
 \be
\Delta x=2\int_\delta^{h_0}\frac{dh}{h'}\,.
 \ee
Implicitly, in both of these expressions, we are assuming that eq.~\reef{bana} allows us to
solve for $h'$ in terms of $h$. However, the details of this solution will be unimportant in the following.
Now combining the two equations above yields
 \be
\sigma=\cL_0\,\Delta x+4\int_\delta^{h_0}\! dh\,\frac{\partial \cL}{\partial P}\, h'\,.
 \labell{43}
 \ee
Differentiating this expression with respect to $h_0$ yields
 \be
\frac{d\sigma}{dh_0}=\cL_0\,\frac{d\Delta x}{dh_0}
+\frac{d\cL_0}{dh_0}\,\Delta x+4\int_\delta^{h_0}\!dh\,\frac{d}{dh_0}\(\frac{\partial \cL}{\partial P} h'\)
\,. \labell{42}
 \ee
Here we can utilize eq.~\eqref{bana} to show
 \beqa
\frac{d}{dh_0}\(2\frac{\partial \cL}{\partial P} h'\)&=&\frac{d}{dh_0}\(\frac{2}{\sqrt{P}}\,
\frac{\partial \cL}{\partial P}\,P\)
\nonumber\\
&=&\frac{1}{\sqrt{P}}\(\frac{d\cL}{dh_0}-\frac{d\cL_0}{dh_0}\)
-\frac{1}{\sqrt{P}}\,\frac{\partial \cL}{\partial P}\,\frac{dP}{dh_0}
\nonumber\\
&=&-\frac{1}{\sqrt{P}}\frac{d\cL_0}{dh_0}+\frac{1}{\sqrt{P}}
\(\frac{d\cL}{dh_0}-\frac{\partial \cL}{\partial P}\frac{dP}{dh_0}\)
\nonumber\\
&=&-\frac{1}{h'}\frac{d\cL_0}{dh_0}\,. \labell{44}
 \eeqa
Let us comment on the vanishing of the bracketed term in the third line: As originally presented in eq.~\reef{38x},
$\cL$ is a function of two quantities, $h$ and $P$. Here, $h$ is simply the integration variable while
$P$ is the implicit solution of eq.~\reef{bana}. Therefore all of the dependence of $\cL$ on $h_0$
comes through the latter, \ie $\frac{d\cL}{dh_0}=\frac{\partial \cL}{\partial P}\frac{dP}{dh_0}$,
and hence the combination appearing in the brackets in the third line vanishes.
In any event, substituting this result into eq.~\eqref{42} yields
 \be
\frac{d\sigma}{dh_0}=\cL_0\,\frac{d\Delta x}{dh_0}\,,
 \ee
which allows us to write
 \be
\frac{d\sigma}{d\Delta x}=\frac{d\sigma}{dh_0}\bigg/\frac{d\Delta x}{dh_0}=\cL_0\,.
 \ee

Now applying the same reasoning as presented above in deriving eq.~\reef{boro}, we
arrive at the following expression for the \name entropy
 \be
E=\ell_2\cdots\ell_{d-1}\,\int_0^{\ell_1}\! dx\,\cL_0\,(1+a')\,.
 \labell{boro4}
 \ee
Again, at first sight, this expression and eq.~\reef{goal77x} are quite different, however,
we will now show that the integrands only differ by a total derivative and hence both
yield the same result.

To begin, recall that the shift $a$ between the tangent point $x$ and
the midpoint of the corresponding interval $x_c(x)$ can be expressed as:
$a=\int_z^{h_0} dh/\partial_{\tilde x}h$, as in eq.~\reef{boro2}. Next,
we devise the analog of the auxiliary function in eq.~\reef{aux1}
 \be
A=2 \int_z^{h_0}\!dh\,\frac{\partial \cL}{\partial P}\, h'\,.
\labell{aux2}
 \ee
Note the similarity between $A$ above and the second term in eq.~\eqref{43}, except that their lower ends of
integration are different. Differentiating this quantity with respect to $x$ and applying eq.~\reef{bana},
one can show
 \be
\frac{dA}{dx}=-(\cL-\cL_0)-\cL_0'\,a\,.
 \labell{iden8}
 \ee
This identity then simplifies the difference between the integrands in eqs.~\reef{goal77x} and \reef{boro4}
to reveal a total derivative,
 \be
\cL_0\,(1+a')-\cL=\cL_0\,a'+\cL_0'\,a+A'=(\cL_0\,a+A)'\,.
 \ee
Hence in this general case, we have once again established the equivalence of the gravitational entropy \reef{goal77x}
in the bulk and the \name entropy \reef{boro4} in the boundary theory.

\section{Discussion}
\labell{discuss}

The spacetime entanglement conjecture of \cite{new1} naturally leads to the question
of whether there are boundary observables corresponding to the Bekenstein-Hawking
entropy of bulk surfaces in the context of the AdS/CFT correspondence. Of course, the
Ryu-Takayanagi prescription \cite{rt1,rt2} provides the first positive response to this question
since it equates $S_\mt{BH}$ of certain extremal surfaces in the bulk with the entanglement
entropy of regions in the boundary theory. Ref.~\cite{hole} made the exciting observation
that $S_\mt{BH}$ evaluated on closed surfaces in AdS$_3$ could be interpreted as
the \name entropy of a family of intervals in the boundary theory. In the present paper,
we have extended this observation in a variety of ways. In particular, we have shown
that the connection between \name entropy in the boundary theory and gravitational entropy
of bulk surfaces extends to higher dimensions, to general holographic backgrounds, and
to higher curvature bulk theories, including Lovelock gravity. Hence this new holographic
equivalence seems to be on quite a robust footing.

Of course, our results only provide the initial steps towards establishing this equivalence in
complete generality and there remain a variety of challenges towards this goal. In  particular,
our analysis assumed planar symmetry, \ie the bulk surfaces had a profile $z=z(x)$ which only
depended on a single (Cartesian) coordinate in the boundary. More generally, one would like
to understand the general situation in higher dimensions where the bulk surface depends on
all of the boundary coordinates. It would seem that in this situation, the relevant \name entropy
would be associated with a tiling the boundary geometry by finite regions. Hence one challenge
would be to establish a systematic approach to constructing such tilings which would allow us
to reconstruct arbitrary profiles $z=z(\vec x)$ in the continuum limit. Of course, another challenge
in this regard would be to construct the equivalent of the \name entropy \reef{residue2} for
such general tilings. The latter is likely to include entanglement entropies of more complicated
intersections and unions of boundary regions and so a technical challenge would be to
explicitly evaluate the holographic entanglement entropy for such complex regions.
Another question would be to establish the equivalence
between \name entropy and gravitational entropy for bulk surfaces, which are not
confined to a constant time slice. Progress on this topic will be reported in \cite{prep}.

Other longer range issues in developing this program would include: One finds quite
generally that there are `barriers' beyond which extremal surfaces will not penetrate
in holographic backgrounds \cite{aron5}, \eg the horizon of a stationary
black hole \cite{aron5,veron5,GBbar}. Hence it is clear that the present approach must be
revised to describe the gravitational entropy of bulk surfaces crossing such barriers.
Another issue arises if one would like to describe the full gravitational entropy in
the bulk beyond the leading large $N$ approximation. As discussed in the context
of holographic entanglement entropy \cite{also}, one should expect quite
generically that there will be corrections to the entanglement at
order $N^0$ which go beyond the usual Bekenstein-Hawking entropy. However, it
seems that the current approach cannot differentiate such entanglement for degrees
of freedom localized on either side of the bulk surface or localized on the same side
of the bulk surface but at still with large separation in the bulk.\footnote{We would
like to thank Juan Maldancena for pointing out this issue.}

We might re-iterate that the original discussion in \cite{hole} related the construction
of a `hole' in the AdS$_3$ spacetime to accelerated observers in the bulk. From this perspective, it is natural
to associate the intervals on the boundary with the corresponding causal wedges \cite{causal0,causal1}
in the bulk. That is, one may consider the \name entropy as constructed using the causal
holographic information associated with the boundary intervals. However, as discussed in section \ref{cause},
this interpretation seems specific to three-dimensional AdS space. In higher dimensions, constructing
a version of the \name entropy \reef{reschi} in this way leads to divergent results. Again, the origin of these divergences
is that beyond the area law contribution,
the boundary divergences appearing in the causal holographic information are nonlocal \cite{ben}
and so these subleading divergences do not cancel in eq.~\reef{reschi}. Of course, one can still consider
the causal development of each of the regions which are used to define the \name entropy in the boundary. These
boundary regions are then naturally associated with a region of the bulk spacetime known as the `entanglement
wedge', using the extremal surface which determines the holographic entanglement entropy \cite{wedge3}.
These entanglement wedges may still play a role in understanding the full significance of differential entropy.

An important feature of the \name entropy is that the boundary strips have an intrinsic ordering and that eq.~\reef{residue}
only involves the entanglement entropy of the intersections of consecutive regions. For example, in the discussion
near the beginning of section \ref{higher}, a given strip will intersect with $2\Delta x n/\ell_1$ other intervals, which
diverges in the continuum limit as $n\to\infty$. However, the \name entropy only considers the intersections of $I_k$
with its two `neighbours' $I_{k\pm 1}$.  An interesting observation made in section \ref{subtle} was that the intrinsic
ordering of the boundary regions does not necessarily correspond to an ordering in the position of the strips
along the boundary, although it does correspond to an ordering in the position along the bulk surface. We also found
that the back-tracking of the boundary intervals, \ie $x_c'(x)<0$, occured when the corresponding extremal surface
in the bulk had a local radius of curvature smaller than that of the bulk surface at the point where these two surfaces are tangent
to one another.

As discussed in section \ref{ads3}, even before taking the continuum limit, the \name entropy of a discrete
family of intervals in the boundary of AdS$_3$ will bound the gravitational entropy of the outer envelope. Of course,
this result also extends to higher dimensions in the situation where there is a planar symmetry and the boundary
is covered by a finite family of strips. Hence for a holographic theory, the \name entropy is generically bounded
below by some finite positive quantity, \ie the gravitational entropy of the dual outer envelope. If instead, we
consider a generic QFT, we can apply strong subadditivity in the same situation to produce an analogous
lower bound corresponding the entanglement entropy of the union of all the strips. However, if the QFT is in a
pure state, this entanglement entropy vanishes and so we can only say that the \name entropy is a positive
(or zero) quantity. Hence the bound for holographic theories seems to be a stronger one. It would be interesting
if more stringent bounds, \ie the \name entropy is greater than some finite quantity, could be established for
generic QFT's using other methods. Alternatively, it may be that these inequalities can be used to establish
a nontrivial test for the behavior of holographic quantum field theories.

An important question which remains is to find a direct interpretation of the \name entropy in terms of
the boundary theory. The proposal put forward in \cite{hole} is as follows: This entropy corresponds to the
maximum entropy of a global state (\ie of a density matrix describing the entire system)
which is consistent with the combined observables measured with the separate density matrices
associated with the individual intervals.\footnote{See \cite{aron} for related discussions in the context
of causal holographic information.} This quantity may be naturally referred to as the `residual
entropy'\footnote{Of course, `residual entropy' is already has a common usage in condensed matter
physics \cite{horse2}.}
or `residual uncertainty' --- \eg see \cite{jandb}.

More pragmatically, we observe that the \name entropy is related
to the derivative of the entanglement entropy with respect to the size of the boundary region --- see also
\cite{prep}.\footnote{Hence our choice of the name: \name entropy.}
That is, in the continuum limit, the discrete differences of entanglement entropies become
derivatives.  In particular, eqs.~\reef{differ9} and \reef{boro4} can be expressed as
 \be
E=\int_0^{\ell_1}\!dx\ \frac{dS}{d\Delta x}\,(1+a')\,.
 \labell{bound88}
 \ee
Now if we set aside the holographic picture, the interpretation of $a(x)$ is not entirely clear in terms of
the boundary theory. However, we must also note that in the
above integral, $x$ refers to the position on the bulk surface for which each interval
is contributing and so in terms of the boundary theory, it is not a natural variable
with which to express the above integral. However, let us recall that
in our construction, $a(x)$ is defined as the displacement from $x$ to the midpoint
of the corresponding interval, \ie $x_c=x+a(x)$ and hence we have
$\frac{\partial x_c}{\partial x}=1+a'$.  Therefore the above integral includes
precisely the Jacobian needed to convert eq.~\reef{bound88} into an
integral over $x_c$,
 \be
E=\sum_{i=1} ^N \ \int_{x_i}^{x_{i+1}}\!dx_c\ \frac{dS}{d\Delta x}\,.
 \labell{bound888}
 \ee
Here, we have introduced the sum in the above expression as a reminder that in general,
$x_c$ has turning points where $\frac{\partial x_c}{\partial x}=0$ -- see section \ref{subtle}.
Labelling these turning points as $x_i$ with $i=1,\cdots,N$ and assuming $x_2>x_1$, we comment
that the terms in the sum with even $i$ are actually making a negative contribution to $E$,
\ie $x_{i+1}<x_i$ when $i$ is even.  Of course, the same sign appears for the corresponding
contributions in eq.~\reef{bound88} since these are the regions where $1+a'<0$. We should
comment that the construction in \cite{hole} refers directly to the analog of this expression
\reef{bound888} for global coordinates.
We also observe that this perspective seems to relate the \name entropy to the `entropy density'
introduced in \cite{density}.

A similar boundary interpretation can be attributed to the geometric formula for the
bulk gravitational entropy. For example, recall eq.~\reef{goal1} for the Bekenstein-Hawking entropy
of a surface described by the profile $z=z(x)$ in AdS$_3$. Using eqs.~\reef{cc3}, \reef{slope}
and \reef{radius}, as well as $x_c=x+a(x)$, this formula can be re-expressed as
 \be
\frac{\A}{4\Gn}=\sum_{i=1} ^N \ \int_{x_i}^{x_{i+1}}\!dx_c\ \frac{dS}{d\Delta x}\
g(a,\Delta x)
\labell{30}
 \ee
where
 \be
g(a,\Delta x)=\frac{1-\partial_{x_c}a}{1-4a^2/\Delta x^2}\,.
 \ee
Note that in this case, the integrals are positive for all of the segments. In particular, the numerator
in $g(a,\Delta x)$, which is equal to $1/(1+a')$, is negative on the segments with even $i$.
Eq.~\reef{30} can also be extended to higher dimensions, however, the definition of
the density $g(a,\Delta x)$ becomes more involved. Again, the interpretation of $a(x)$ in terms of
the boundary theory remains unclear. Setting this issue aside, it would be interesting if one could establish
that eqs.~\reef{bound88} and \reef{30} yield the same result without referring to holography.

\vskip 1cm

\section*{Acknowledgements}

We would like to thank Vijay Balasubramanian, Nikolay Bobev, Jan de Boer,
Horacio Casini, Bartek Czech, Netta Engelhardt, Masafumi Fukuma,
Veronika Hubeny, Juan Maldacena, Mukund Rangamani, Cobi Sonnenschein
and James Sully for useful comments and discussions.
We also thank Nikolay Bobev for comments
on a preliminary draft of this paper.
Research at Perimeter Institute is supported by the
Government of Canada through Industry Canada and by the Province of Ontario
through the Ministry of Research \& Innovation. RCM also acknowledges support
from an NSERC Discovery grant and funding from the Canadian Institute for
Advanced Research. RCM also thanks the Kavli Institute for Theoretical Physics
for hospitality during the final stages of this project. Research at the KITP was
supported in part by the National Science Foundation under Grant No.~NSF
PHY11-25915.
SS thanks Perimeter's Visiting Graduate Fellows Program and the people at
the Perimeter Institute for their kind hospitality,
where most this work was done. SS also acknowledges the Bilateral International Exchange Program
of Kyoto University, which supported in part his visit to Perimeter Institute.

\appendix

\section{Entropy functional for Lovelock gravity}
\labell{loveA}

In this section, we examine the gravitational entropy functional for Lovelock
gravity \cite{lovel}. In particular, we show that for holographic background geometries of the form
given in eq.~\reef{back}, if this entropy is evaluated on a bulk surface with a profile $z=z(x)$,
then the resulting functional takes the form given in eq.~\reef{goal77x}.

The general action for Lovelock
gravity \cite{lovel} in $d+1$ dimensions can be written as
 \beq
I = \frac{1}{2\lp^{d-1}} \int {d}^{d+1}x \, \sqrt{-g}\, \left[
\frac{d(d-1)}{L^2} + R +
\sum_{p=2}^{\left\lfloor\frac{d+1}{2}\right\rfloor}
c_p\,L^{2p-2}\,\cL_{2p}(R) \right]\,,
 \labell{GBactg}
 \eeq
where $\left\lfloor\frac{d+1}{2}\right\rfloor$ denotes the integer part
of $(d+1)/2$ and $c_p$ are dimensionless coupling constants for the
higher curvature terms. These higher order interactions are defined as
 \beq
\cL_{2p}(R) \equiv \frac{1}{2^p}\ \delta_{\mu_1\,\mu_2\,\cdots\,
\mu_{2p-1}\,\mu_{2p}}^{\nu_1\,\nu_2\,\cdots\, \nu_{2p-1}\,\nu_{2p}}\
R^{\mu_1\mu_2}{}_{\nu_1\nu_2}\,\cdots\,
R^{\mu_{2p-1}\mu_{2p}}{}_{\nu_{2p-1}\nu_{2p}}\,,
 \labell{term0}
 \eeq
which is proportional to the Euler density on a 2$p$-dimensional
manifold. Here, we are using $\delta_{\mu_1\,\mu_2\,\cdots\,
\mu_{2p-1}\,\mu_{2p}}^{\nu_1\,\nu_2\,\cdots\, \nu_{2p-1}\,\nu_{2p}}$ to
denote the totally antisymmetric product of $2p$ Kronecker delta
symbols. Of course, the cosmological constant and Einstein terms could
be incorporated into the sum as $\cL_0$ and $\cL_2$, respectively.
However, we exhibit them explicitly above to establish our
normalization for the Planck length, as well as the length scale $L$.

The original motivation to study this theory \reef{GBactg} was that
the resulting equations of motion are second order in derivatives
\cite{lovel}. However recently, there has been renewed interest in these theories in the
context of the AdS/CFT correspondence. In particular, these theories
provide toy models where the central charges in the boundary CFT
are different from one another \cite{gb2,highc,highc2}.
These theories also proved useful in
discussions of holographic hydrodynamics and the consistency of the
boundary CFT \cite{EtasGB,gb2}, as well as of holographic
$c$-theorems \cite{cosmic,gb}.

Black hole entropy in the Lovelock theories was first
discussed in \cite{tedJ}, where using a Hamiltonian
approach, the following expression was derived
 \beq
S_\mt{JM} = \frac{2 \pi}{\lp^{d-1}} \int
d^{d-1}x\,\sqrt{h}\,\left[
1+\sum_{p=2}^{\left\lfloor\frac{d+1}{2}\right\rfloor}
p\,c_p\,L^{2p-2}\,\cL_{2p-2}(\R) \right]\,.
 \labell{Waldformula3}
 \eeq
Here $\R^{\al\bt}{}_{\ga\de}$ are the components of the intrinsic
curvature tensor of the slice of the event horizon on which this
expression is evaluated. We should note that this expression differs
from the standard Wald entropy \cite{wald1} by terms involving the extrinsic
curvature of the surface. Hence the two formulae will agree when
evaluating the horizon entropy for a stationary black hole with a
Killing horizon.

Now in studying holographic entanglement entropy for Lovelock gravity,
it was argued that the correct extension of eq.~\reef{define} was to
simply replace the BH entropy by eq.~\reef{Waldformula3}.
This prescription was shown to pass various nontrivial consistency tests
involving the universal contribution to the entanglement entropy for even
dimensional boundary theories \cite{highc,highc2}. However, we should add
that the recent derivation of the RT prescription \cite{aitor}
can be extended to derive this new prescription for Lovelock gravity \cite{xifriend}.

We now turn to evaluating $S_\mt{JM}$ on a surface with a profile $z=z(x)$ in a
background geometry of the form described by eq.~\reef{back}.
First, the induced metric on the surface can be written as
 \be
ds^2=g_1\,(1+f(z)\,z'^2)\,dx^2+\sum_{i=2}^{d-1}g_i\,(dx^i)^2\,,
 \ee
where the $g_i$'s are all functions of $z$ only.  With a bit of work, the components
of the Riemann tensor for this metric can be determined as
 \beqa
\mathcal{R}^{xi}{}_{xi}&=&\frac{1}{2g_ig_1\sqrt{Q}}\[-\(\frac{g_i'}{\sqrt{Q}}\)'
+\frac{1}{2}\(\frac{g_1'}{g_1}+\frac{g_i'}{g_i}\)\frac{g_i'}{\sqrt{Q}}\]\,,
\nonumber\\
\mathcal{R}^{kl}{}_{kl}&=&-\frac{1}{4}\frac{g_k'}{g_k}\frac{g_i'}{g_l}\frac{1}{g_1Q}\,,
\labell{curvation}
 \eeqa
where we have defined $Q=1+f(z)\,z'^2$. Now the only potentially problematic
contributions proportional to $z''$ come from the term with $(g_i'/\sqrt{Q})'$
in $\mathcal{R}^{xi}{}_{xi}$. However, a key feature of $S_\mt{JM}$ is that the curvature
contributions take the same form as in eq.~\reef{term0}. Hence $\mathcal{R}^{xi}{}_{xi}$ will
appear at most once in any of these expressions. In particular, if we focus on the potentially
problematic terms, we have
\beq
\sqrt{h}\cL_{2p-2}({\cal R})\propto \sqrt{h}\,\mathcal{R}^{xi}{}_{xi}\,\mathcal{R}^{kl}_{\ph1\ph1kl}\cdots
\mathcal{R}^{mn}_{\ph1\ph1mn}
\labell{rest}
\eeq
where there are $p-2$ curvatures beyond the factor of $\mathcal{R}^{xi}{}_{xi}$. Now gathering
up all of the factors of $z'$ and using $z'^2= (Q-1)/f(z)$, we may write these potentially
problematic terms as
 \beqa
\sqrt{h}\,\mathcal{R}^{xi}_{\ph1\ph1xi}\,\mathcal{R}^{kl}_{\ph1\ph1kl}\cdots\mathcal{R}^{mn}_{\ph1\ph1mn}
&\simeq& F(z)\,\(\frac{Q-1}{Q}\)^{p-2}\,\(\frac{\sqrt{Q-1}}{\sqrt{Q}}\)'
= \frac{F(z)}{2p-1}\,\[\(\frac{Q-1}{Q}\)^{p-\frac12}\]'
\nonumber\\
&= & \frac{1}{2p-1}\,\[F(z)\(\frac{Q-1}{Q}\)^{p-\frac12}\]'
-\frac{\partial_zF(z)\,z'}{2p-1}\,\(\frac{Q-1}{Q}\)^{p-\frac12}
\nonumber\\
&=&-\frac{\partial_zF(z)}{(2p-1)\,\sqrt{f(z)}}\ \frac{(Q-1)^p}{Q^{p-\frac12}}+\cdots
\,.\labell{cardf}
 \eeqa
Hence the potentially problematic terms are eliminated by integrating by parts.
Further, we note that odd powers of $z'\propto\sqrt{Q-1}$ are avoided in the
final expression. Therefore
the integrand of generalized entropy functional \reef{Waldformula3} for the Lovelock gravity
takes the desired form given in eq.~\eqref{goal77x}.


\end{document}